\documentclass[11pt]{article}
\usepackage{amsmath,amssymb}
\usepackage{array}
\newcount\figureno     \figureno=0
\newdimen\figdim       \figdim=70mm
\def\figureinc{%
   \global\advance\figureno by 1%
}
\def\figcaption#1#2#3{\hbox to #2{\hss{\vbox{\hsize=#2 \parindent=0pt
        {\bf Figure \number\figureno#3 :\ }#1}}\hss}
}

\begin{document}
\baselineskip 100pt
\renewcommand{\baselinestretch}{1.5}
\renewcommand{\arraystretch}{0.666666666}
{\large
\parskip.2in
\numberwithin{equation}{section}
\newcommand{\be}{\begin{equation}}
\newcommand{\ee}{\end{equation}}
\newcommand{\ben}{\begin{equation*}}
\newcommand{\een}{\end{equation*}}
\newcommand{\eqalinb}{\begin{eqnarray}}
\newcommand{\eqaline}{\end{eqnarray}}
\newcommand{\br}{\bar}
\newcommand{\fr}{\frac}
\newcommand{\lm}{\lambda}
\newcommand{\ra}{\rightarrow}
\newcommand{\al}{\alpha}
\newcommand{\bt}{\beta}
\newcommand{\z}{\zeta}
\newcommand{\pa}{\partial}
\newcommand{\hs}{\hspace{5mm}}
\newcommand{\up}{\upsilon}
\newcommand{\bigb}{\hspace{7mm}}
\newcommand{\dg}{\dagger}
\newcommand{\vphi}{\vec{\varphi}}
\newcommand{\ve}{\varepsilon}
\newcommand{\acc}{\\[3mm]}
\newcommand{\dl}{\delta}
\newcommand{\sdil}{\ensuremath{\rlap{\raisebox{.15ex}{$\mskip
6.5mu\scriptstyle+ $}}\subset}}
\newcommand{\sdir}{\ensuremath{\rlap{\raisebox{.15ex}{$\mskip
6.5mu\scriptstyle+ $}}\supset}}
\def\tablecap#1{\vskip 3mm \centerline{#1}\vskip 5mm}
\def\p#1{\partial_#1}
\newcommand{\pd}[2]{\frac{\partial #1}{\partial #2}}
\newcommand{\pdn}[3]{\frac{\partial #1^{#3}}{\partial #2^{#3}}}
\def\DP#1#2{D_{#1}\varphi^{#2}}
\def\dP#1#2{\partial_{#1}\varphi^{#2}}
\def\xh{\hat x}
\newcommand{\Ref}[1]{(\ref{#1})}
\def\ld{\,\ldots\,}

\def\C{{\mathbb C}}
\def\Z{{\mathbb Z}}
\def\R{{\mathbb R}}
\def\mod#1{ \vert #1 \vert }
\def\chapter#1{\hbox{Introduction.}}
\def\Sin{\hbox{sin}}
\def\Cos{\hbox{cos}}
\def\Exp{\hbox{exp}}
\def\Ln{\hbox{ln}}
\def\Tan{\hbox{tan}}
\def\Cot{\hbox{cot}}
\def\Sinh{\hbox{sinh}}
\def\Cosh{\hbox{cosh}}
\def\Tanh{\hbox{tanh}}
\def\Asin{\hbox{asin}}
\def\Acos{\hbox{acos}}
\def\Atan{\hbox{atan}}
\def\Asinh{\hbox{asinh}}
\def\Acosh{\hbox{acosh}}
\def\Atanh{\hbox{atanh}}
\def\frac#1#2{{\textstyle{#1\over #2}}}

\def\ph{\varphi_{m,n}}
\def\phl{\varphi_{m-1,n}}
\def\phr{\varphi_{m+1,n}}
\def\varphil{\varphi_{m-1,n}}
\def\varphir{\varphi_{m+1,n}}
\def\varphit{\varphi_{m,n+1}}
\def\varphib{\varphi_{m,n-1}}
\def\pht{\varphi_{m,n+1}}
\def\phb{\varphi_{m,n-1}}
\def\phbl{\varphi_{m-1,n-1}}
\def\phbr{\varphi_{m+1,n-1}}
\def\phtl{\varphi_{m-1,n+1}}
\def\phtr{\varphi_{m+1,n+1}}
\def\u{u_{m,n}}
\def\ul{u_{m-1,n}}
\def\ur{u_{m+1,n}}
\def\ut{u_{m,n+1}}
\def\ub{u_{m,n-1}}
\def\utr{u_{m+1,n+1}}
\def\ubl{u_{m-1,n-1}}
\def\utl{u_{m-1,n+1}}
\def\ubr{u_{m+1,n-1}}
\def\v{v_{m,n}}
\def\vl{v_{m-1,n}}
\def\vr{v_{m+1,n}}
\def\vt{v_{m,n+1}}
\def\vb{v_{m,n-1}}
\def\vtr{v_{m+1,n+1}}
\def\vbl{v_{m-1,n-1}}
\def\vtl{v_{m-1,n+1}}
\def\vbr{v_{m+1,n-1}}

\def\U{U_{m,n}}
\def\Ul{U_{m-1,n}}
\def\Ur{U_{m+1,n}}
\def\Ut{U_{m,n+1}}
\def\Ub{U_{m,n-1}}
\def\Utr{U_{m+1,n+1}}
\def\Ubl{U_{m-1,n-1}}
\def\Utl{U_{m-1,n+1}}
\def\Ubr{U_{m+1,n-1}}
\def\V{V_{m,n}}
\def\Vl{V_{m-1,n}}
\def\Vr{V_{m+1,n}}
\def\Vt{V_{m,n+1}}
\def\Vb{V_{m,n-1}}
\def\Vtr{V_{m+1,n+1}}
\def\Vbl{V_{m-1,n-1}}
\def\Vtl{V_{m-1,n+1}}
\def\Vbr{V_{m+1,n-1}}
\def\tr{{\rm tr}\,}

\def\a{\alpha}
\def\b{\beta}
\def\g{\gamma}
\def\d{\delta}
\def\ep{\epsilon}
\def\e{\varepsilon}
\def\z{\zeta}
\def\t{\theta}
\def\k{\kappa}
\def\l{\lambda}
\def\s{\sigma}
\def\f{\varphi}
\def\w{\omega}
\def\v{{\hbox{v}}}
\def\u{{\hbox{u}}}
\def\x{{\hbox{x}}}

\newcommand{\ie}{{\it i.e.}}
\newcommand{\cmod}[1]{ \vert #1 \vert ^2 }
\newcommand{\cmodn}[2]{ \vert #1 \vert ^{#2} }
\newcommand{\nhat}{\mbox{\boldmath$\hat n$}}
\nopagebreak[3]
\bigskip

\title{ \bf Supersymmetric formulation of the minimal surface equation: algebraic aspects}
\vskip 1cm

\bigskip
\author{
A.~M. Grundland\thanks{email address: grundlan@crm.umontreal.ca}
\\
Centre de Recherches Math{\'e}matiques, Universit{\'e} de Montr{\'e}al,\\
C. P. 6128, Succ.\ Centre-ville, Montr{\'e}al, (QC) H3C 3J7,
Canada\\ Universit\'{e} du Qu\'{e}bec, Trois-Rivi\`{e}res, CP500 (QC) G9A 5H7, Canada \acc A. J. Hariton\thanks{email
address: hariton@crm.umontreal.ca}
\\
Centre de Recherches Math{\'e}matiques, Universit{\'e} de Montr{\'e}al,\\
C. P. 6128, Succ.\ Centre-ville, Montr{\'e}al, (QC) H3C 3J7,
Canada} \date{}

\maketitle
\begin{abstract}
In this paper, a supersymmetric extension of the minimal surface equation is formulated. Based on this formulation, a Lie superalgebra of infinitesimal symmetries of this equation is determined. A classification of the one-dimensional subalgebras is performed, which results in a list of 143 conjugacy classes with respect to action by the supergroup generated by the Lie superalgebra. The symmetry reduction method is used to obtain invariant solutions of the supersymmetric minimal surface equation. The classical minimal surface equation is also examined and its group-theoretical properties are compared with those of the supersymmetric version.\\
\end{abstract}





\vspace{3mm}


\newpage

\section{Introduction}

In recent years, there has been a considerable amount of interest in supersymmetric (SUSY) models involving odd (anticommuting) Grassmann variables and superalgebras. Supersymmetry was introduced in the theory of elementary particles and their interactions and forms an essential component of attempts to obtain a unification of all physical forces \cite{Aitchison}. A number of supersymmetric extensions have been formulated for both classical and quantum mechanical systems. In particular, such supersymmetric generalizations have been constructed for hydrodynamic-type systems (e.g. the Korteweg-de Vries equation \cite{Mathieu,Labelle,LiuManas,BarcelosNeto,HusAyaWin}, the Sawada--Kotera equation \cite{TianLiu}, polytropic gas dynamics \cite{Das,Polytropic} and a Gaussian irrotational compressible fluid \cite{Gaussian}) as well as other nonlinear wave equations (e.g. the Schr\"{o}dinger equation \cite{Unterberger} and the sine/sinh-Gordon equation \cite{Grammaticos,Gomes,SSG,SKG}. Parametrizations of strings and Nambu-Goto membranes have been used to supersymmetrize the Chaplygin gas in $(1+1)$ and $(2+1)$ dimensions respectively \cite{Jackiw,Bergner,Polychronakos}. In addition, it was proposed that non-Abelian fluid mechanics and color magnetohydrodynamics could be used to describe a quark-gluon plasma \cite{Jackiw,Pi}.

\noindent In this paper, we formulate a supersymmetric extension of the minimal surface equation and investigate its group-theoretical properties. The concept of minimal surfaces was originally devised by Joseph-Louis Lagrange in the mid-eighteenth century \cite{Bianchi} and still remains an active subject of research and applications. We consider a smooth orientable conformally parametrized surface ${\mathcal F}$ defined by the immersion $\vec{F}:{\mathcal R}\rightarrow \mathbb{R}^3$ of a complex domain ${\mathcal R}\subset\mathbb{C}$ into three-dimensional Euclidean space $\mathbb{R}^3$. We consider a variation of ${\mathcal F}$ along a vector field $\vec{v}$ which vanishes on the boundary of ${\mathcal F}$: $\vec{v}\mid_{\partial{\mathcal F}}=0$. The corresponding variation of the area of ${\mathcal F}$ in the small parameter $\varepsilon$ (where $\varepsilon<<1$) is, up to higher terms in $\varepsilon$,
\begin{equation}
A(\vec{F}+\varepsilon\vec{v})-A(\vec{F})=-2\varepsilon\int_{\mathcal F} \vec{v}\cdot\vec{H} dA+\ldots,
\label{intr01}
\end{equation}
where $\vec{H}$ is the mean curvature vector on ${\mathcal F}$. Surfaces with vanishing mean curvature ($\vec{H}=0$) are called minimal surfaces. The conformal metric associated with the surface ${\mathcal F}$ is $\Omega=e^{u}dzd{\bar{z}}$, where $z$ and $\bar{z}$ are coordinates on ${\mathcal R}$ and $u$ is a real-valued function of $z$ and $\bar{z}$. If we re-label the variables $z$ and $\bar{z}$ as $x$ and $y$ respectively, then the real-valued function $u$ satisfies the partial differential equation (PDE) \cite{Spivak}
\begin{equation}
(1+(u_x)^2)u_{yy}-2u_xu_yu_{xy}+(1+(u_y)^2)u_{xx}=0,
\label{minimals}
\end{equation}
which is called the minimal surface (MS) equation. Equation (\ref{minimals}) can be written in the form of the following conservation law:
\begin{equation}
\partial_x\left({u_x\over \sqrt{1+(u_x)^2+(u_y)^2}}\right)+\partial_y\left({u_y\over \sqrt{1+(u_x)^2+(u_y)^2}}\right)=0,
\label{CL}
\end{equation}
which can be derived from the variational principle for the Lagrangian density
\begin{equation}
{\cal L}=\sqrt{1+(u_x)^2+(u_y)^2}.
\label{Lag}
\end{equation}
A conformally parametrized surface is minimal if and only if it can be locally expressed as the graph of a solution of equation (\ref{minimals}). The minimal surface and its related equations appear in many areas of physics and mathematics, such as fluid dynamics \cite{Lamb,Roz}, continuum mechanics \cite{Chandrasekhar,Hill}, nonlinear field theory \cite{Boilat,Nelson,David}, plasma physics \cite{Jackiw,Luban,Chen}, nonlinear optics \cite{Sommerfeld,Luneburg} and the theory of fluid membranes \cite{Davidov,Ou,Safran}. Using the Wick rotation $y=it$, one can transform equation (\ref{minimals}) to the scalar Born-Infeld equation \cite{Arik,Whitham}
\begin{equation}
(1+(u_x)^2)u_{tt}-2u_xu_tu_{xt}-(1-(u_t)^2)u_{xx}=0.
\label{Born}
\end{equation}

\noindent In this paper, we construct a supersymmetric extension of the minimal surface equation (\ref{minimals}) using a superspace and superfield formalism. The space $\{(x,y)\}$ of independent variables is extended to the superspace $\{(x,y,\theta_1,\theta_2)\}$ while 
the bosonic surface function $u(x,y)$ is replaced by the bosonic superfield $\Phi(x,y,\theta_1,\theta_2)$ defined in terms of bosonic and fermionic-valued fields of $x$ and $y$. Following the construction of our supersymmetric extension, we determine a Lie superalgebra of infinitesimal symmetries of our extended equation. We then classify the one-dimensional subalgebras of this Lie superalgebra into conjugation classes with respect to action by the Lie supergroup generated by the Lie superalgebra, and we use the symmetry reduction method to obtain invariant solutions of the SUSY equation. The advantage of using such group-theoretical methods to analyze our supersymmetrized equation is that these methods are systematic and involve regular algorithms which, in theory, can be used without having to make additional assumptions. Finally, we revisit and expand the group-theoretical analysis of the classical minimal surface equation and compare the obtained results to those found for the supersymmetric extension of the minimal surface equation.

\section{Supersymmetric version of the minimal surface equation}

Grassmann variables are elements of a Grassmann algebra $\Lambda$ involving a finite number of Grassmann generators $\zeta_1,\zeta_2,\ldots,\zeta_k$ which obey the rules
\begin{equation}
\begin{split}
&\zeta_i\zeta_j=-\zeta_j\zeta_i\mbox{ if }i\neq j,\\
&\zeta_i^2=0\mbox{ for all }i.
\end{split}
\label{grasscond}
\end{equation}
The Grassmann algebra can be decomposed into even and odd parts: $\Lambda=\Lambda_{even}+\Lambda_{odd}$, where $\Lambda_{even}$ consists of all terms involving the product of an even number of generators $1,\zeta_1\zeta_2,\zeta_1\zeta_3,\ldots,\zeta_1\zeta_2\zeta_3\zeta_4,\ldots$, while $\Lambda_{odd}$ consists of all terms involving the product of an odd number of generators $\zeta_1,\zeta_2,\zeta_3,\ldots,\zeta_1\zeta_2\zeta_3,\zeta_1\zeta_2\zeta_4,\ldots$. A Grassmann variable $\kappa$ is called even (or bosonic) if it is a linear combination of terms involving an even number of generators, while it is called odd (or fermionic) if it is a linear combination of terms involving an odd number of generators.\\\\
We now construct a Grassmann-valued extension of the minimal surface equation (\ref{minimals}). The space of independent variables, $\{(x,y)\}$, is extended to a superspace $\{(x,y,\theta_1,\theta_2)\}$ involving two fermionic Grassmann-valued variables $\theta_1$ and $\theta_2$. Also, the bosonic function $u(x,y)$ is generalized to a bosonic-valued superfield $\Phi$ defined as
\begin{equation}
\Phi(x,y,\theta_1,\theta_2)=v(x,y)+\theta_1\phi(x,y)+\theta_2\psi(x,y)+\theta_1\theta_2u(x,y),
\label{superfield}
\end{equation}
where $v(x,y)$ is a bosonic-valued field while $\phi(x,y)$ and $\psi(x,y)$ are fermionic-valued fields. We construct our extension in such a way that it is invariant under the supersymmetry transformations
\begin{equation}
x\longrightarrow x-\underline{\eta_1}\theta_1,\hspace{5mm}\theta_1\longrightarrow \theta_1+\underline{\eta_1},
\label{trQ1}
\end{equation}
and
\begin{equation}
y\longrightarrow y-\underline{\eta_2}\theta_2,\hspace{5mm}\theta_2\longrightarrow \theta_2+\underline{\eta_2},
\label{trQ2}
\end{equation}
where $\underline{\eta_1}$ and $\underline{\eta_2}$ are odd-valued parameters. Throughout this paper, we use the convention that underlined constants are fermionic-valued. The transformations (\ref{trQ1}) and (\ref{trQ2}) are generated by the infinitesimal supersymmetry generators
\begin{equation}
Q_1=\partial_{\theta_1}-\theta_1\partial_{x}\hspace{1cm}\mbox{and}\hspace{1cm}Q_2=\partial_{\theta_2}-\theta_2\partial_{y},
\label{supersymmetry}
\end{equation}
respectively. These generators satisfy the anticommutation relations
\begin{equation}
\{Q_1,Q_1\}=-2\partial_x,\hspace{1cm}\{Q_2,Q_2\}=-2\partial_y,\hspace{1cm}\{Q_1,Q_2\}=0.
\label{anticommutators}
\end{equation}
To make the superfield model invariant under the transformations generated by $Q_1$ and $Q_2$, we construct the equation in terms of the following covariant derivatives:
\begin{equation}
D_1=\partial_{\theta_1}+\theta_1\partial_{x}\hspace{1cm}\mbox{and}\hspace{1cm}D_2=\partial_{\theta_2}+\theta_2\partial_{y}.
\label{covariant}
\end{equation}
These covariant derivative operators possess the following properties
\begin{equation}
\begin{split}
& D_1^2=\partial_x,\hspace{1cm}D_2^2=\partial_y,\hspace{1cm}\{D_1,D_2\}=0,\hspace{1cm}\{D_1,Q_1\}=0,\hspace{1cm}\\ & \{D_1,Q_2\}=0,\hspace{1cm}\{D_2,Q_1\}=0,\hspace{1cm}\{D_2,Q_2\}=0.
\end{split}
\label{covprop}
\end{equation}
Combining different covariant derivatives $D_1^m$ and $D_2^n$ of the superfield $\Phi$ of various orders, where $m$ and $n$ are positive integers, we obtain the most general form of the supersymmetric extension of equation (\ref{minimals}). Since this expression is very involved, we instead present the following sub-case as our superymmetric extension of the MS equation, and will refer to it as such. We obtain the equation
\begin{equation}
\begin{split}
&D_2^4\Phi+(D_1^2\Phi)(D_1^3D_2\Phi)(D_1D_2^5\Phi)-2(D_1^2\Phi)(D_1D_2^3\Phi)(D_1^3D_2^3\Phi)+D_1^4\Phi\\ &+(D_2^2\Phi)(D_1D_2^3\Phi)(D_1^5D_2\Phi)=0.
\label{eqmotion1}
\end{split}
\end{equation}
In terms of derivatives with respect to $x$, $y$, $\theta_1$ and $\theta_2$, equation (\ref{eqmotion1}) can be written in the form
\begin{equation}
\begin{split}
& \Phi_{yy}+\Phi_{xx}\\ &+\Phi_{x}(-\Phi_{x\theta_1\theta_2}+\theta_1\Phi_{xx\theta_2}-\theta_2\Phi_{xy\theta_1}+\theta_1\theta_2\Phi_{xxy})\times\\ & \hspace{5mm} (-\Phi_{yy}\theta_1\theta_2+\theta_1\Phi_{xyy\theta_2}-\theta_2\Phi_{yyy\theta_1}+\theta_1\theta_2\Phi_{xyyy})\\
& -2\Phi_{x}(-\Phi_{y\theta_1\theta_2}+\theta_1\Phi_{xy\theta_2}-\theta_2\Phi_{yy\theta_1}+\theta_1\theta_2\Phi_{xyy})\times\\ & \hspace{5mm} (-\Phi_{xy\theta_1\theta_2}+\theta_1\Phi_{xxy\theta_2}-\theta_2\Phi_{xyy\theta_1}+\theta_1\theta_2\Phi_{xxyy})\\
&+\Phi_{y}(-\Phi_{y\theta_1\theta_2}+\theta_1\Phi_{xy\theta_2}-\theta_2\Phi_{yy\theta_1}+\theta_1\theta_2\Phi_{xyy})\times\\ & \hspace{5mm} (-\Phi_{xx\theta_1\theta_2}+\theta_1\Phi_{xxx\theta_2}-\theta_2\Phi_{xxy\theta_1}+\theta_1\theta_2\Phi_{xxxy})\\ &=0.
\end{split}
\label{eqmotion2}
\end{equation}
In what follows, we will refer to equation (\ref{eqmotion2}) as the supersymmetric minimal surface equation (SUSY MS equation).\\\\
The partial derivatives satisfy the generalized Leibniz rule
\begin{equation}
\partial_{\theta_i}(fg)=(\partial_{\theta_i}f)g+(-1)^{\mbox{deg}(f)}f(\partial_{\theta_i}g),
\end{equation}
if $\theta_i$ is a fermionic variable and we define
\begin{equation}
\mbox{deg}(f)=\begin{cases}
               0 & \mbox{ if } f \mbox{ is even},\\
               1 & \mbox{ if } f \mbox{ is odd}.
              \end{cases}
\end{equation}
The partial derivatives with respect to the odd coordinates satisfy $\partial_{\theta_i}(\theta_j)=\delta_j^i$, where the indices $i$ and $j$ each stand for 1 or 2 and $\delta_j^i$ is the Kronecker delta function. The operators $\partial_{\theta_1}$, $\partial_{\theta_2}$, $Q_1$, $Q_2$, $D_1$ and $D_2$ change the parity of a bosonic function to that of a fermionic function and vice-versa.\\\\
When dealing with higher-order derivatives, the symbol $f_{x_1x_2x_3\ldots x_{k-1}x_k}$ denotes the derivative $\partial_{x_k}\partial_{x_{k-1}}\ldots\partial_{x_3}\partial_{x_2}\partial_{x_1}(f)$ where the order must be preserved for the sake of consistency.
Throughout this paper, we use the convention that if $f(g(x))$ is a composite function, then
\begin{equation}
\dfrac{\partial f}{\partial x}=\dfrac{\partial g}{\partial x}\cdot\dfrac{\partial f}{\partial g}.
\label{composite}
\end{equation}
The interchange of mixed derivatives with proper respect for the ordering of odd variables is assumed throughout. For a review of recent developments in this subject see e.g. Freed \cite{Freed} and Varadarajan \cite{Varadarajan}.

\section{Lie symmetries of the supersymmetric minimal surface equation}

A symmetry supergroup $G$ of a supersymmetric system is a local supergroup of transformations acting on the Cartesian product of submanifolds ${\mathcal X}\times{\mathcal U}$, where ${\mathcal X}$ is the space of independent variables $\{(x,y,\theta_1,\theta_2)\}$ and ${\mathcal U}$ is the space of dependent superfields, which in this case involves only the superfield $\Phi$. In order to find symmetries of the SUSY MS equation, we make use of the theory described in the book by Olver \cite{Olver} in order to determine superalgebras of infinitesimal symmetries.\\\\
In order to determine the Lie point superalgebra of infinitesimal symmetries, we look for a bosonic vector field of the form
\begin{equation}
\begin{split}
{\mathbf v}=&\xi_1(x,y,\theta_1,\theta_2)\partial_x+\xi_2(x,y,\theta_1,\theta_2)\partial_y+\rho_1(x,y,\theta_1,\theta_2)\partial_{\theta_1}\\ &+\rho_2(x,y,\theta_1,\theta_2)\partial_{\theta_2}+\Lambda(x,y,\theta_1,\theta_2)\partial_{\Phi},
\label{vectorfield}
\end{split}
\end{equation}
where $\xi_1$, $\xi_2$ and $\Lambda$ are bosonic-valued functions, while $\rho_1$ and $\rho_2$ are fermionic-valued functions. The prolongation formulas allowing us to find the symmetries are very involved and will not be presented here. Moreover, it should be noted that the symmetry criterion has not yet been conclusively demonstrated for the case of equations involving Grassmann variables.\\\\
The following infinitesimal transformations were found to be symmetry generators of the SUSY MS equation
\begin{equation}
\begin{split}
& P_1=\partial_x,\hspace{1cm}P_2=\partial_y,\hspace{1cm}P_3=\partial_{\theta_1},\hspace{1cm}P_4=\partial_{\theta_2},\hspace{1cm}P_5=\partial_{\Phi},\hspace{1cm}\\ & D=2x\partial_x+2y\partial_y+\theta_1\partial_{\theta_1}+\theta_2\partial_{\theta_2}+4\Phi\partial_{\Phi}, \\ & Q_1=\partial_{\theta_1}-\theta_1\partial_{x},\hspace{1cm} Q_2=\partial_{\theta_2}-\theta_2\partial_{y}.
\end{split}
\label{symmetries}
\end{equation}
These eight generators span a Lie superalgebra ${\mathcal G}$ of infinitesimal symmetries of the SUSY MS equation. Here, $P_1$, $P_2$, $P_3$ and $P_4$ generate translations in the $x$, $y$, $\theta_1$ and $\theta_2$ directions respectively, while $P_5$ generates a shift in the superfield $\Phi$. The vector field $D$ corresponds to a dilation involving both bosonic and fermionic variables as well as the superfield $\Phi$. Finally, the fermionic vector fields $Q_1$ and $Q_2$ are simply the supersymmetry transformations identified in (\ref{supersymmetry}). The supercommutation relations involving the generators of the superalgebra ${\mathcal G}$ are listed in Table 1. 

\begin{table}[htbp]
\caption{Supercommutation table for the Lie superalgebra ${\mathcal G}$ generated by the vector fields (\ref{symmetries}). Here, for each pair of generators $X$ and $Y$, we calculate either the commutator $[X,Y]=XY-YX$ if either $X$ or $Y$ are bosonic, or the anticommutator $\{X,Y\}=XY+YX$ if both $X$ and $Y$ are fermionic.}
\begin{center}
\begin{tabular}{|c||c|c|c|c|c|c|c|c|}\hline
 & $\mathbf{D}$ & $\mathbf{P_1}$ & $\mathbf{P_3}$ & $\mathbf{Q_1}$ & $\mathbf{P_2}$ & $\mathbf{P_4}$ & $\mathbf{Q_2}$ & $\mathbf{P_5}$ \\\hline\hline
$\mathbf{D}$ & $0$ & $-2P_1$ & $-P_3$ & $-Q_1$ & $-2P_2$ & $-P_4$ & $-Q_2$ & $-4P_5$ \\\hline
$\mathbf{P_1}$ & $2P_1$ & $0$ & $0$ & $0$ & $0$ & $0$ & $0$ & $0$ \\\hline
$\mathbf{P_3}$ & $P_3$ & $0$ & $0$ & $-P_1$ & $0$ & $0$ & $0$ & $0$ \\\hline
$\mathbf{Q_1}$ & $Q_1$ & $0$ & $-P_1$ & $-2P_1$ & $0$ & $0$ & $0$ & $0$ \\\hline
$\mathbf{P_2}$ & $2P_2$ & $0$ & $0$ & $0$ & $0$ & $0$ & $0$ & $0$ \\\hline
$\mathbf{P_4}$ & $P_4$ & $0$ & $0$ & $0$ & $0$ & $0$ & $-P_2$ & $0$ \\\hline
$\mathbf{Q_2}$ & $Q_2$ & $0$ & $0$ & $0$ & $0$ & $-P_2$ & $-2P_2$ & $0$ \\\hline
$\mathbf{P_5}$ & $4P_5$ & $0$ & $0$ & $0$ & $0$ & $0$ & $0$ & $0$ \\\hline
\end{tabular}
\end{center}
\end{table}

\noindent The Lie superalgebra ${\mathcal G}$ can be decomposed into the following combination of semidirect and direct sums:
\begin{equation}
{\mathcal G}=\{D\}\sdir\{\{P_1,P_3,Q_1\}\oplus\{P_2,P_4,Q_2\}\oplus\{P_5\}\}.
\label{decomposed}
\end{equation}
It should be noted that the symmetries found for the SUSY MS equation (\ref{eqmotion2}) are qualitatively different from those found previously for the SUSY version of the equations of confomally parametrized surfaces with non-zero mean curvature \cite{Bertrand}.

\section{Classification of Subalgebras for the Lie Superalgebra}

We proceed to classify the one-dimensional Lie subalgebras of the superalgebra ${\mathcal G}$ generated by (\ref{symmetries}) into conjugacy classes under the action of the Lie supergroup $G=\mbox{exp}({\mathcal G})$ generated by ${\mathcal G}$. We construct our list of representative subalgebras in such a way that each one-dimensional subalgebra of ${\mathcal G}$ is conjugate to one and only one element of the list. Such a classification is useful because subalgebras that are conjugate to each other lead to invariant solutions that are equivalent in the sense that one can be transformed to the other by a suitable symmetry. Therefore, it is not necessary to perform symmetry reduction on two different subalgebras that are conjugate to each other.\\\\
In order to classify the Lie superalgebra ${\mathcal G}$ given in (\ref{decomposed}) we make use of the procedures given in \cite{Winternitz}. In what follows, $\alpha$, $r$, $k$ and $\ell$ are bosonic constants, $\underline{\mu}$, $\underline{\nu}$, $\underline{\eta}$, $\underline{\lambda}$, $\underline{\rho}$ and $\underline{\sigma}$ are fermionic constants, and $\varepsilon=\pm 1$. We begin by considering the subalgebra ${\mathcal S}_1=\{P_1,P_3,Q_1\}$. Consider a general element of ${\mathcal S}_1$ which can be written as the linear combination $X=\alpha P_1+\underline{\mu}P_3+\underline{\nu}Q_1$ and examine how this element changes under the action of the one-parameter group generated by the generator: $Y=rP_1+\underline{\eta}P_3+\underline{\lambda}Q_1$. This action is performed through the Baker-Campbell-Hausdorff formula
\begin{equation}
X\longrightarrow \mbox{Ad}_{\mbox{exp}(Y)}X=X+[Y,X]+\frac{1}{2\!}[Y,[Y,X]]+\ldots+\frac{1}{3\!}[Y,[Y,[Y,X]]]+\ldots
\label{BCHformula}
\end{equation}
We obtain
\begin{equation}
\begin{split}
[Y,X]&=[rP_1+\underline{\eta}P_3+\underline{\lambda}Q_1,\alpha P_1+\underline{\mu}P_3+\underline{\nu}Q_1]\\
     &=[\underline{\eta}P_3,\underline{\nu}Q_1]+[\underline{\lambda}Q_1,\underline{\mu}P_3]+[{\lambda}Q_1,\underline{\nu}Q_1]\\
     &=(\underline{\eta}\underline{\nu}+\underline{\lambda}\underline{\mu}+2\underline{\lambda}\underline{\nu})P_1,
\end{split}
\end{equation}
\begin{equation}
[Y[Y,X]]=[rP_1+\underline{\eta}P_3+\underline{\lambda}Q_1,(\underline{\eta}\underline{\nu}+\underline{\lambda}\underline{\mu}+2\underline{\lambda}\underline{\nu})P_1]=0.
\end{equation}
So we have
\begin{equation}
\{\alpha P_1+\underline{\mu}P_3+\underline{\nu}Q_1\}\longrightarrow \{(\alpha+\underline{\eta}\underline{\nu}+\underline{\lambda}\underline{\mu}+2\underline{\lambda}\underline{\nu}) P_1+\underline{\mu}P_3+\underline{\nu}Q_1\}.
\end{equation}
Therefore, aside from a change in the $P_1$ coefficient, each element of the form $\{\alpha P_1+\underline{\mu}P_3+\underline{\nu}Q_1\}$ is conjugate only to itself. This gives us the subalgebras
\begin{equation}
\begin{split}
&{\mathcal G}_1=\{P_1\}, \hspace{5mm} {\mathcal G}_2=\{\underline{\mu}P_3\}, \hspace{5mm}
{\mathcal G}_3=\{\underline{\mu}Q_1\}, \hspace{5mm}  {\mathcal G}_4=\{P_1+\underline{\mu}P_3\}, \\ &
{\mathcal G}_5=\{P_1+\underline{\mu}Q_1\}, \hspace{5mm} {\mathcal G}_6=\{\underline{\mu}P_3+\underline{\nu}Q_1\}, \hspace{5mm}
{\mathcal G}_7=\{P_1+\underline{\mu}P_3+\underline{\nu}Q_1\}.
\end{split}
\label{list1}
\end{equation}
An analogous classification is performed for the subalgebra ${\mathcal S}_2=\{P_2,P_4,Q_2\}$, from where we obtain the subalgebras
\begin{equation}
\begin{split}
&{\mathcal G}_8=\{P_2\},  \hspace{5mm}
{\mathcal G}_9=\{\underline{\mu}P_4\}, \hspace{5mm} {\mathcal G}_{10}=\{\underline{\mu}Q_2\}, \hspace{5mm} 
{\mathcal G}_{11}=\{P_2+\underline{\mu}P_4\}, \\  
& {\mathcal G}_{12}=\{P_2+\underline{\mu}Q_2\}, \hspace{5mm} {\mathcal G}_{13}=\{\underline{\mu}P_4+\underline{\nu}Q_2\}, \hspace{5mm} {\mathcal G}_{14}=\{P_2+\underline{\mu}P_4+\underline{\nu}Q_2\}.
\end{split}
\label{list2}
\end{equation}
\\
The next step is to classify the direct sum of the algebras ${\mathcal S}_1$ and ${\mathcal S}_2$, that is, to classify
\begin{equation}
{\mathcal S}={\mathcal S}_1\oplus {\mathcal S}_2=\{P_1,P_3,Q_1\}\oplus \{P_2,P_4,Q_2\},
\end{equation}
using the Goursat method of subalgebra classification \cite{Goursat1,Goursat2}. Each non-twisted subalgebra of ${\mathcal S}$ is constructed by selecting one subalgebra of ${\mathcal S}_1$ and finding its direct sum with a subalgebra of ${\mathcal S}_2$. The non-twisted one-dimensional subalgebras of ${\mathcal S}$ are the combined subalgebras ${\mathcal G}_{1}$ to ${\mathcal G}_{14}$ listed in (\ref{list1}) and (\ref{list2}). The twisted subalgebras of ${\mathcal S}$ are formed as follows. If $A\in {\mathcal S}_1$ and $B\in {\mathcal S}_2$, then $A$ and $B$ can be twisted together if there exists a homomorphism from $A$ to $B$, say $\tau(A)=B$. The twisted subalgebra is then obtained by taking $\{A+\tau(A)\}$. This results in the additional subalgebras
\begin{displaymath}
\begin{split}
&{\mathcal G}_{15}=\{P_1+kP_2\}, \hspace{5mm} {\mathcal G}_{16}=\{P_1+\underline{\mu}P_4\}, \hspace{5mm}
{\mathcal G}_{17}=\{P_1+\underline{\mu}Q_2\}, \\ & {\mathcal G}_{18}=\{P_1+kP_2+\underline{\mu}P_4\}, \hspace{5mm}
{\mathcal G}_{19}=\{P_1+kP_2+\underline{\mu}Q_2\}, \hspace{5mm} {\mathcal G}_{20}=\{P_1+\underline{\mu}P_4+\underline{\nu}Q_2\}, \\ &
{\mathcal G}_{21}=\{P_1+kP_2+\underline{\mu}P_4+\underline{\nu}Q_2\}, \hspace{5mm} {\mathcal G}_{22}=\{P_2+\underline{\mu}P_3\}, \hspace{5mm}
{\mathcal G}_{23}=\{\underline{\mu}P_3+\underline{\nu}P_4\}, \\ & {\mathcal G}_{24}=\{\underline{\mu}P_3+\underline{\nu}Q_2\}, \hspace{5mm}
{\mathcal G}_{25}=\{P_2+\underline{\mu}P_3+\underline{\nu}P_4\}, \hspace{5mm} {\mathcal G}_{26}=\{P_2+\underline{\mu}P_3+\underline{\nu}Q_2\}, \\ &
{\mathcal G}_{27}=\{\underline{\mu}P_3+\underline{\nu}P_4+\underline{\rho}Q_2\}, \hspace{5mm} {\mathcal G}_{28}=\{P_2+\underline{\mu}P_3+\underline{\nu}P_4+\underline{\rho}Q_2\}, \hspace{5mm}
{\mathcal G}_{29}=\{P_2+\underline{\mu}Q_1\}, \hspace{1cm} \\ & {\mathcal G}_{30}=\{\underline{\mu}P_4+\underline{\nu}Q_1\}, \hspace{5mm}
{\mathcal G}_{31}=\{\underline{\mu}Q_1+\underline{\nu}Q_2\}, \hspace{5mm} {\mathcal G}_{32}=\{P_2+\underline{\mu}P_4+\underline{\nu}Q_1\}, \\ &
{\mathcal G}_{33}=\{P_2+\underline{\mu}Q_1+\underline{\nu}Q_2\}, \hspace{5mm} {\mathcal G}_{34}=\{\underline{\mu}P_4+\underline{\nu}Q_1+\underline{\rho}Q_2\}, \\ &
{\mathcal G}_{35}=\{P_2+\underline{\mu}P_4+\underline{\nu}Q_1+\underline{\rho}Q_2\}, \hspace{5mm} {\mathcal G}_{36}=\{P_1+kP_2+\underline{\mu}P_3\}, \\ &
{\mathcal G}_{37}=\{P_1+\underline{\mu}P_3+\underline{\nu}P_4\}, \hspace{5mm} {\mathcal G}_{38}=\{P_1+\underline{\mu}P_3+\underline{\nu}Q_2\}, \\ &
{\mathcal G}_{39}=\{P_1+kP_2+\underline{\mu}P_3+\underline{\nu}P_4\}, \hspace{5mm} {\mathcal G}_{40}=\{P_1+kP_2+\underline{\mu}P_3+\underline{\nu}Q_2\}, \\ &
\end{split}
\end{displaymath}
\begin{equation}
\begin{split}
&{\mathcal G}_{41}=\{P_1+\underline{\mu}P_3+\underline{\nu}P_4+\underline{\rho}Q_2\}, \hspace{5mm} {\mathcal G}_{42}=\{P_1+kP_2+\underline{\mu}P_3+\underline{\nu}P_4+\underline{\rho}Q_2\}, \\ &
{\mathcal G}_{43}=\{P_1+kP_2+\underline{\mu}Q_1\}, \hspace{5mm} {\mathcal G}_{44}=\{P_1+\underline{\mu}P_4+\underline{\nu}Q_1\}, \\ &
{\mathcal G}_{45}=\{P_1+\underline{\mu}Q_1+\underline{\nu}Q_2\}, \hspace{5mm} {\mathcal G}_{46}=\{P_1+kP_2+\underline{\mu}P_4+\underline{\nu}Q_1\}, \\ &
{\mathcal G}_{47}=\{P_1+kP_2+\underline{\mu}Q_1+\underline{\nu}Q_2\}, \hspace{5mm} {\mathcal G}_{48}=\{P_1+\underline{\mu}P_4+\underline{\nu}Q_1+\underline{\rho}Q_2\}, \\ &
{\mathcal G}_{49}=\{P_1+kP_2+\underline{\mu}P_4+\underline{\nu}Q_1+\underline{\rho}Q_2\}, \hspace{5mm} {\mathcal G}_{50}=\{P_2+\underline{\mu}P_3+\underline{\nu}Q_1\}, \\ &
{\mathcal G}_{51}=\{\underline{\mu}P_3+\underline{\nu}P_4+\underline{\rho}Q_1\}, \hspace{5mm} {\mathcal G}_{52}=\{\underline{\mu}P_3+\underline{\nu}Q_1+\underline{\rho}Q_2\}, \\ &
{\mathcal G}_{53}=\{P_2+\underline{\mu}P_3+\underline{\nu}P_4+\underline{\rho}Q_1\}, \hspace{5mm} {\mathcal G}_{54}=\{P_2+\underline{\mu}P_3+\underline{\nu}Q_1+\underline{\rho}Q_2\}, \\ &
{\mathcal G}_{55}=\{\underline{\mu}P_3+\underline{\nu}P_4+\underline{\rho}Q_1+\underline{\sigma}Q_2\}, \hspace{5mm} {\mathcal G}_{56}=\{P_2+\underline{\mu}P_3+\underline{\nu}P_4+\underline{\rho}Q_1+\underline{\sigma}Q_2\}, \\ &
{\mathcal G}_{57}=\{P_1+kP_2+\underline{\mu}P_3+\underline{\nu}Q_1\}, \hspace{5mm} {\mathcal G}_{58}=\{P_1+\underline{\mu}P_3+\underline{\nu}P_4+\underline{\rho}Q_1\}, \\ &
{\mathcal G}_{59}=\{P_1+\underline{\mu}P_3+\underline{\nu}Q_1+\underline{\rho}Q_2\}, \hspace{5mm} {\mathcal G}_{60}=\{P_1+kP_2+\underline{\mu}P_3+\underline{\nu}P_4+\underline{\rho}Q_1\}, \\ &
{\mathcal G}_{61}=\{P_1+kP_2+\underline{\mu}P_3+\underline{\nu}Q_1+\underline{\rho}Q_2\}, \hspace{5mm} {\mathcal G}_{62}=\{P_1+\underline{\mu}P_3+\underline{\nu}P_4+\underline{\rho}Q_1+\underline{\sigma}Q_2\}, \\ &
{\mathcal G}_{63}=\{P_1+kP_2+\underline{\mu}P_3+\underline{\nu}P_4+\underline{\rho}Q_1+\underline{\sigma}Q_2\}.
\end{split}
\label{list3}
\end{equation}

\noindent Next, we consider the one-dimensional subalgebras of the semi-direct sum 
\begin{equation}
\tilde{\mathcal S}={\mathcal S}\oplus\{P_5\}=\{\{P_1,P_3,Q_1\}\oplus \{P_2,P_4,Q_2\}\}\oplus\{P_5\},
\end{equation}
Using the Goursat method as described above, we obtain, in addition to the subalgebras already listed in (\ref{list1}), (\ref{list2}) and (\ref{list3}), the subalgebras
\begin{displaymath}
\begin{split}
&{\mathcal G}_{64}=\{P_5\}, \hspace{5mm} {\mathcal G}_{65}=\{P_1+kP_5\}, \hspace{5mm} {\mathcal G}_{66}=\{P_5+\underline{\mu}P_3\}, \hspace{5mm}
{\mathcal G}_{67}=\{P_5+\underline{\mu}Q_1\}, \hspace{5cm} \\ &  {\mathcal G}_{68}=\{P_1+kP_5+\underline{\mu}P_3\}, \hspace{5mm}
{\mathcal G}_{69}=\{P_1+kP_5+\underline{\mu}Q_1\}, \\ & {\mathcal G}_{70}=\{P_5+\underline{\mu}P_3+\underline{\nu}Q_1\}, \hspace{5mm}
{\mathcal G}_{71}=\{P_1+kP_5+\underline{\mu}P_3+\underline{\nu}Q_1\}, \\ & {\mathcal G}_{72}=\{P_2+kP_5\},  \hspace{5mm}
{\mathcal G}_{73}=\{P_5+\underline{\mu}P_4\}, \hspace{5mm} {\mathcal G}_{74}=\{P_5+\underline{\mu}Q_2\}, \\ &
{\mathcal G}_{75}=\{P_2+kP_5+\underline{\mu}P_4\}, \hspace{5mm}
{\mathcal G}_{76}=\{P_2+kP_5+\underline{\mu}Q_2\}, \\ & {\mathcal G}_{77}=\{P_5+\underline{\mu}P_4+\underline{\nu}Q_2\}, \hspace{5mm} {\mathcal G}_{78}=\{P_2+kP_5+\underline{\mu}P_4+\underline{\nu}Q_2\},\\ & 
{\mathcal G}_{79}=\{P_1+kP_2+\ell P_5\}, \hspace{5mm} {\mathcal G}_{80}=\{P_1+kP_5+\underline{\mu}P_4\}, \\ &
{\mathcal G}_{81}=\{P_1+kP_5+\underline{\mu}Q_2\}, \hspace{5mm} {\mathcal G}_{82}=\{P_1+kP_2+\ell P_5+\underline{\mu}P_4\}, \\ &
{\mathcal G}_{83}=\{P_1+kP_2+\ell P_5+\underline{\mu}Q_2\}, \hspace{5mm} {\mathcal G}_{84}=\{P_1+kP_5+\underline{\mu}P_4+\underline{\nu}Q_2\}, \\ & {\mathcal G}_{85}=\{P_1+kP_2+\ell P_5+\underline{\mu}P_4+\underline{\nu}Q_2\}, \hspace{5mm} {\mathcal G}_{86}=\{P_2+kP_5+\underline{\mu}P_3\}, \\ &
{\mathcal G}_{87}=\{P_5+\underline{\mu}P_3+\underline{\nu}P_4\}, \hspace{5mm} {\mathcal G}_{88}=\{P_5+\underline{\mu}P_3+\underline{\nu}Q_2\},
\end{split}
\end{displaymath}
\begin{equation}
\begin{split}
&{\mathcal G}_{89}=\{P_2+kP_5+\underline{\mu}P_3+\underline{\nu}P_4\}, \hspace{5mm} {\mathcal G}_{90}=\{P_2+kP_5+\underline{\mu}P_3+\underline{\nu}Q_2\}, \\ &
{\mathcal G}_{91}=\{P_5+\underline{\mu}P_3+\underline{\nu}P_4+\underline{\rho}Q_2\}, \hspace{5mm} {\mathcal G}_{92}=\{P_2+kP_5+\underline{\mu}P_3+\underline{\nu}P_4+\underline{\rho}Q_2\}, \\ &
{\mathcal G}_{93}=\{P_2+kP_5+\underline{\mu}Q_1\}, \hspace{5mm} {\mathcal G}_{94}=\{P_5+\underline{\mu}P_4+\underline{\nu}Q_1\}, \\ &
{\mathcal G}_{95}=\{P_5+\underline{\mu}Q_1+\underline{\nu}Q_2\}, \hspace{5mm} {\mathcal G}_{96}=\{P_2+kP_5+\underline{\mu}P_4+\underline{\nu}Q_1\}, \\ &
{\mathcal G}_{97}=\{P_2+kP_5+\underline{\mu}Q_1+\underline{\nu}Q_2\}, \hspace{5mm} {\mathcal G}_{98}=\{P_5+\underline{\mu}P_4+\underline{\nu}Q_1+\underline{\rho}Q_2\}, \\ &
{\mathcal G}_{99}=\{P_2+kP_5+\underline{\mu}P_4+\underline{\nu}Q_1+\underline{\rho}Q_2\}, \hspace{5mm} {\mathcal G}_{100}=\{P_1+kP_2+\ell P_5+\underline{\mu}P_3\}, \\ &
{\mathcal G}_{101}=\{P_1+kP_5+\underline{\mu}P_3+\underline{\nu}P_4\}, \hspace{5mm} {\mathcal G}_{102}=\{P_1+kP_5+\underline{\mu}P_3+\underline{\nu}Q_2\}, \\ &
{\mathcal G}_{103}=\{P_1+kP_2+\ell P_5+\underline{\mu}P_3+\underline{\nu}P_4\}, \\ & {\mathcal G}_{104}=\{P_1+kP_2+\ell P_5+\underline{\mu}P_3+\underline{\nu}Q_2\}, \\ &
{\mathcal G}_{105}=\{P_1+kP_5+\underline{\mu}P_3+\underline{\nu}P_4+\underline{\rho}Q_2\}, \\ & {\mathcal G}_{106}=\{P_1+kP_2+\ell P_5+\underline{\mu}P_3+\underline{\nu}P_4+\underline{\rho}Q_2\}, \\ &
{\mathcal G}_{107}=\{P_1+kP_2+\ell P_5+\underline{\mu}Q_1\}, \hspace{5mm} {\mathcal G}_{108}=\{P_1+kP_5+\underline{\mu}P_4+\underline{\nu}Q_1\}, \\ &
{\mathcal G}_{109}=\{P_1+kP_5+\underline{\mu}Q_1+\underline{\nu}Q_2\}, \hspace{5mm} {\mathcal G}_{110}=\{P_1+kP_2+\ell P_5+\underline{\mu}P_4+\underline{\nu}Q_1\}, \hspace{1.1cm} \\ &
{\mathcal G}_{111}=\{P_1+kP_2+\ell P_5+\underline{\mu}Q_1+\underline{\nu}Q_2\}, \\ & {\mathcal G}_{112}=\{P_1+kP_5+\underline{\mu}P_4+\underline{\nu}Q_1+\underline{\rho}Q_2\}, \\ &
{\mathcal G}_{113}=\{P_1+kP_2+\ell P_5+\underline{\mu}P_4+\underline{\nu}Q_1+\underline{\rho}Q_2\}, \\ & {\mathcal G}_{114}=\{P_2+kP_5+\underline{\mu}P_3+\underline{\nu}Q_1\}, \hspace{5mm}
{\mathcal G}_{115}=\{P_5+\underline{\mu}P_3+\underline{\nu}P_4+\underline{\rho}Q_1\}, \\ & {\mathcal G}_{116}=\{P_5+\underline{\mu}P_3+\underline{\nu}Q_1+\underline{\rho}Q_2\}, \hspace{5mm}
{\mathcal G}_{117}=\{P_2+kP_5+\underline{\mu}P_3+\underline{\nu}P_4+\underline{\rho}Q_1\}, \\ & {\mathcal G}_{118}=\{P_2+kP_5+\underline{\mu}P_3+\underline{\nu}Q_1+\underline{\rho}Q_2\}, \\ &
{\mathcal G}_{119}=\{P_5+\underline{\mu}P_3+\underline{\nu}P_4+\underline{\rho}Q_1+\underline{\sigma}Q_2\}, \\ & {\mathcal G}_{120}=\{P_2+kP_5+\underline{\mu}P_3+\underline{\nu}P_4+\underline{\rho}Q_1+\underline{\sigma}Q_2\}, \\ &
{\mathcal G}_{121}=\{P_1+kP_2+\ell P_5+\underline{\mu}P_3+\underline{\nu}Q_1\}, \\ & {\mathcal G}_{122}=\{P_1+kP_5+\underline{\mu}P_3+\underline{\nu}P_4+\underline{\rho}Q_1\}, \\ &
{\mathcal G}_{123}=\{P_1+kP_5+\underline{\mu}P_3+\underline{\nu}Q_1+\underline{\rho}Q_2\}, \\ & {\mathcal G}_{124}=\{P_1+kP_2+\ell P_5+\underline{\mu}P_3+\underline{\nu}P_4+\underline{\rho}Q_1\}, \\ &
{\mathcal G}_{125}=\{P_1+kP_2+\ell P_5+\underline{\mu}P_3+\underline{\nu}Q_1+\underline{\rho}Q_2\}, \\ & {\mathcal G}_{126}=\{P_1+kP_5+\underline{\mu}P_3+\underline{\nu}P_4+\underline{\rho}Q_1+\underline{\sigma}Q_2\},  \\ &
{\mathcal G}_{127}=\{P_1+kP_2+\ell P_5+\underline{\mu}P_3+\underline{\nu}P_4+\underline{\rho}Q_1+\underline{\sigma}Q_2\}.
\end{split}
\label{list4}
\end{equation}

\ \\
Finally, we classify the complete semidirect sum superalgebra ${\mathcal G}=\{D\}\sdir\tilde{\mathcal S}$ using the method of splitting and non-splitting subalgebras \cite{Splitting1,Splitting2}. The splitting subalgebras of ${\mathcal G}$ are formed by combining the dilation $\{D\}$ or the trivial element $\{0\}$ with each of the subalgebras of $\tilde{\mathcal S}$ in a semidirect sum of the form $F\sdir N$, where $F=\{D\}$ or $F=\{0\}$ and $N$ is a subalgebra of the classification of $\tilde{\mathcal S}$.  The splitting one-dimensional subalgebras of ${\mathcal G}$ are the combined subalgebras ${\mathcal G}_{1}$ to ${\mathcal G}_{127}$ listed in (\ref{list1}), (\ref{list2}), (\ref{list3}) and (\ref{list4}) together with the subalgebra ${\mathcal G}_{128}=\{D\}$. For non-splitting subalgebras, we consider spaces of the form
\begin{equation}
V=\{D+\sum\limits_{i=1}^{s}c_iZ_i\},
\end{equation}
where the $Z_i$ form a basis of ${\mathcal S}$. The resulting possibilities are further classified by observing which are conjugate to each other under the action of the complete group generated by ${\mathcal G}$. This analysis provides us with the additional subalgebras
\begin{displaymath}
\begin{split}
&{\mathcal G}_{129}=\{D+\varepsilon P_1\}, \hspace{5mm} {\mathcal G}_{130}=\{D+\underline{\mu}P_3\},  \hspace{5mm}
{\mathcal G}_{131}=\{D+\underline{\mu}Q_1\}, \\ &{\mathcal G}_{132}=\{D+\varepsilon P_1+\underline{\mu}P_3\},  \hspace{5mm}
{\mathcal G}_{133}=\{D+\varepsilon P_1+\underline{\mu}Q_1\}, \\ &{\mathcal G}_{134}=\{D+\underline{\mu}P_3+\underline{\nu}Q_1\},  \hspace{5mm}
{\mathcal G}_{135}=\{D+\varepsilon P_1+\underline{\mu}P_3+\underline{\nu}Q_1\}, \\ & {\mathcal G}_{136}=\{D+\varepsilon P_2\},  \hspace{5mm}
{\mathcal G}_{137}=\{D+\underline{\mu}P_4\}, \hspace{5mm} {\mathcal G}_{138}=\{D+\underline{\mu}Q_2\},  \\ &
{\mathcal G}_{139}=\{D+\varepsilon P_2+\underline{\mu}P_4\}, \hspace{5mm} {\mathcal G}_{140}=\{D+\varepsilon P_2+\underline{\mu}Q_2\},  \\ &
{\mathcal G}_{141}=\{D+\underline{\mu}P_4+\underline{\nu}Q_2\}, \hspace{5mm} {\mathcal G}_{142}=\{D+\varepsilon P_2+\underline{\mu}P_4+\underline{\nu}Q_2\},  \\ &
{\mathcal G}_{143}=\{D+\varepsilon P_1+kP_2\}, \hspace{5mm} {\mathcal G}_{144}=\{D+\varepsilon P_1+\underline{\mu}P_4\},  \\ &
{\mathcal G}_{145}=\{D+\varepsilon P_1+\underline{\mu}Q_2\}, \hspace{5mm} {\mathcal G}_{146}=\{D+\varepsilon P_1+kP_2+\underline{\mu}P_4\}, \\ &
{\mathcal G}_{147}=\{D+\varepsilon P_1+kP_2+\underline{\mu}Q_2\}, \hspace{5mm} {\mathcal G}_{148}=\{D+\varepsilon P_1+\underline{\mu}P_4+\underline{\nu}Q_2\}, \\ &
{\mathcal G}_{149}=\{D+\varepsilon P_1+kP_2+\underline{\mu}P_4+\underline{\nu}Q_2\}, \hspace{5mm} {\mathcal G}_{150}=\{D+\varepsilon P_2+\underline{\mu}P_3\},  \\ &
{\mathcal G}_{151}=\{D+\underline{\mu}P_3+\underline{\nu}P_4\}, \hspace{5mm} {\mathcal G}_{152}=\{D+\underline{\mu}P_3+\underline{\nu}Q_2\}, \\ &
{\mathcal G}_{153}=\{D+\varepsilon P_2+\underline{\mu}P_3+\underline{\nu}P_4\}, \hspace{5mm} {\mathcal G}_{154}=\{D+\varepsilon P_2+\underline{\mu}P_3+\underline{\nu}Q_2\},  \\ &
{\mathcal G}_{155}=\{D+\underline{\mu}P_3+\underline{\nu}P_4+\underline{\rho}Q_2\}, \hspace{5mm} {\mathcal G}_{156}=\{D+\varepsilon P_2+\underline{\mu}P_3+\underline{\nu}P_4+\underline{\rho}Q_2\}, \hspace{5cm}  \\ &
{\mathcal G}_{157}=\{D+\varepsilon P_2+\underline{\mu}Q_1\}, \hspace{5mm} {\mathcal G}_{158}=\{D+\underline{\mu}P_4+\underline{\nu}Q_1\}, \\ &
{\mathcal G}_{159}=\{D+\underline{\mu}Q_1+\underline{\nu}Q_2\}, \hspace{5mm} {\mathcal G}_{160}=\{D+\varepsilon P_2+\underline{\mu}P_4+\underline{\nu}Q_1\}, \\ &
{\mathcal G}_{161}=\{D+\varepsilon P_2+\underline{\mu}Q_1+\underline{\nu}Q_2\}, \hspace{5mm} {\mathcal G}_{162}=\{D+\underline{\mu}P_4+\underline{\nu}Q_1+\underline{\rho}Q_2\},  \\ &
{\mathcal G}_{163}=\{D+\varepsilon P_2+\underline{\mu}P_4+\underline{\nu}Q_1+\underline{\rho}Q_2\}, \hspace{5mm} {\mathcal G}_{164}=\{D+\varepsilon P_1+kP_2+\underline{\mu}P_3\},  
\end{split}
\end{displaymath}
\begin{displaymath}
\begin{split}
&{\mathcal G}_{165}=\{D+\varepsilon P_1+\underline{\mu}P_3+\underline{\nu}P_4\}, \hspace{5mm} {\mathcal G}_{166}=\{D+\varepsilon P_1+\underline{\mu}P_3+\underline{\nu}Q_2\},  \\ &
{\mathcal G}_{167}=\{D+\varepsilon P_1+kP_2+\underline{\mu}P_3+\underline{\nu}P_4\}, \hspace{5mm} {\mathcal G}_{168}=\{D+\varepsilon P_1+kP_2+\underline{\mu}P_3+\underline{\nu}Q_2\},  \\ &
{\mathcal G}_{169}=\{D+\varepsilon P_1+\underline{\mu}P_3+\underline{\nu}P_4+\underline{\rho}Q_2\}, \\ & {\mathcal G}_{170}=\{D+\varepsilon P_1+kP_2+\underline{\mu}P_3+\underline{\nu}P_4+\underline{\rho}Q_2\},  \hspace{5mm}
{\mathcal G}_{171}=\{D+\varepsilon P_1+kP_2+\underline{\mu}Q_1\}, \\ & {\mathcal G}_{172}=\{D+\varepsilon P_1+\underline{\mu}P_4+\underline{\nu}Q_1\},  \hspace{5mm}
{\mathcal G}_{173}=\{D+\varepsilon P_1+\underline{\mu}Q_1+\underline{\nu}Q_2\}, \\ & {\mathcal G}_{174}=\{D+\varepsilon P_1+kP_2+\underline{\mu}P_4+\underline{\nu}Q_1\},  \hspace{5mm}
{\mathcal G}_{175}=\{D+\varepsilon P_1+kP_2+\underline{\mu}Q_1+\underline{\nu}Q_2\}, \\ & {\mathcal G}_{176}=\{D+\varepsilon P_1+\underline{\mu}P_4+\underline{\nu}Q_1+\underline{\rho}Q_2\},  \\ &
{\mathcal G}_{177}=\{D+\varepsilon P_1+kP_2+\underline{\mu}P_4+\underline{\nu}Q_1+\underline{\rho}Q_2\}, \hspace{5mm} {\mathcal G}_{178}=\{D+\varepsilon P_2+\underline{\mu}P_3+\underline{\nu}Q_1\},  \\ &
{\mathcal G}_{179}=\{D+\underline{\mu}P_3+\underline{\nu}P_4+\underline{\rho}Q_1\}, \hspace{5mm} {\mathcal G}_{180}=\{D+\underline{\mu}P_3+\underline{\nu}Q_1+\underline{\rho}Q_2\},  \\ &
{\mathcal G}_{181}=\{D+\varepsilon P_2+\underline{\mu}P_3+\underline{\nu}P_4+\underline{\rho}Q_1\}, \hspace{5mm} {\mathcal G}_{182}=\{D+\varepsilon P_2+\underline{\mu}P_3+\underline{\nu}Q_1+\underline{\rho}Q_2\},  \\ &
{\mathcal G}_{183}=\{D+\underline{\mu}P_3+\underline{\nu}P_4+\underline{\rho}Q_1+\underline{\sigma}Q_2\}, \\ & {\mathcal G}_{184}=\{D+\varepsilon P_2+\underline{\mu}P_3+\underline{\nu}P_4+\underline{\rho}Q_1+\underline{\sigma}Q_2\},  \\ &
{\mathcal G}_{185}=\{D+\varepsilon P_1+kP_2+\underline{\mu}P_3+\underline{\nu}Q_1\}, \hspace{5mm} {\mathcal G}_{186}=\{D+\varepsilon P_1+\underline{\mu}P_3+\underline{\nu}P_4+\underline{\rho}Q_1\}, \hspace{5cm} \\ &
{\mathcal G}_{187}=\{D+\varepsilon P_1+\underline{\mu}P_3+\underline{\nu}Q_1+\underline{\rho}Q_2\}, \\ & {\mathcal G}_{188}=\{D+\varepsilon P_1+kP_2+\underline{\mu}P_3+\underline{\nu}P_4+\underline{\rho}Q_1\},  \\ &
{\mathcal G}_{189}=\{D+\varepsilon P_1+kP_2+\underline{\mu}P_3+\underline{\nu}Q_1+\underline{\rho}Q_2\}, \\ & {\mathcal G}_{190}=\{D+\varepsilon P_1+\underline{\mu}P_3+\underline{\nu}P_4+\underline{\rho}Q_1+\underline{\sigma}Q_2\},  \\ &
{\mathcal G}_{191}=\{D+\varepsilon P_1+kP_2+\underline{\mu}P_3+\underline{\nu}P_4+\underline{\rho}Q_1+\underline{\sigma}Q_2\}, \hspace{5mm}
{\mathcal G}_{192}=\{D+\varepsilon P_5\}, \\ & {\mathcal G}_{193}=\{D+\varepsilon P_1+kP_5\}, \hspace{5mm} {\mathcal G}_{194}=\{D+\varepsilon P_5+\underline{\mu}P_3\}, \\ &
{\mathcal G}_{195}=\{D+\varepsilon P_5+\underline{\mu}Q_1\}, \hspace{5mm}  {\mathcal G}_{196}=\{D+\varepsilon P_1+kP_5+\underline{\mu}P_3\}, \\ &
{\mathcal G}_{197}=\{D+\varepsilon P_1+kP_5+\underline{\mu}Q_1\}, \hspace{5mm} {\mathcal G}_{198}=\{D+\varepsilon P_5+\underline{\mu}P_3+\underline{\nu}Q_1\}, \\ &
{\mathcal G}_{199}=\{D+\varepsilon P_1+kP_5+\underline{\mu}P_3+\underline{\nu}Q_1\}, \hspace{5mm} {\mathcal G}_{200}=\{D+\varepsilon P_2+kP_5\},  \\ &
{\mathcal G}_{201}=\{D+\varepsilon P_5+\underline{\mu}P_4\}, \hspace{5mm} {\mathcal G}_{202}=\{D+\varepsilon P_5+\underline{\mu}Q_2\}, \\ &
{\mathcal G}_{203}=\{D+\varepsilon P_2+kP_5+\underline{\mu}P_4\}, \hspace{5mm}
{\mathcal G}_{204}=\{D+\varepsilon P_2+kP_5+\underline{\mu}Q_2\}, \\ & {\mathcal G}_{205}=\{D+\varepsilon P_5+\underline{\mu}P_4+\underline{\nu}Q_2\}, \hspace{5mm} {\mathcal G}_{206}=\{D+\varepsilon P_2+kP_5+\underline{\mu}P_4+\underline{\nu}Q_2\},\\ & 
{\mathcal G}_{207}=\{D+\varepsilon P_1+kP_2+\ell P_5\}, \hspace{5mm} {\mathcal G}_{208}=\{D+\varepsilon P_1+kP_5+\underline{\mu}P_4\}, \\ &
{\mathcal G}_{209}=\{D+\varepsilon P_1+kP_5+\underline{\mu}Q_2\}, \hspace{5mm} {\mathcal G}_{210}=\{D+\varepsilon P_1+kP_2+\ell P_5+\underline{\mu}P_4\}, \\ &
{\mathcal G}_{211}=\{D+\varepsilon P_1+kP_2+\ell P_5+\underline{\mu}Q_2\}, \hspace{5mm} {\mathcal G}_{212}=\{D+\varepsilon P_1+kP_5+\underline{\mu}P_4+\underline{\nu}Q_2\}, \\ & {\mathcal G}_{213}=\{D+\varepsilon P_1+kP_2+\ell P_5+\underline{\mu}P_4+\underline{\nu}Q_2\}, \hspace{5mm} {\mathcal G}_{214}=\{D+\varepsilon P_2+kP_5+\underline{\mu}P_3\},
\end{split}
\end{displaymath}
\begin{displaymath}
\begin{split}
&{\mathcal G}_{215}=\{D+\varepsilon P_5+\underline{\mu}P_3+\underline{\nu}P_4\}, \hspace{5mm} {\mathcal G}_{216}=\{D+\varepsilon P_5+\underline{\mu}P_3+\underline{\nu}Q_2\}, \\ &
{\mathcal G}_{217}=\{D+\varepsilon P_2+kP_5+\underline{\mu}P_3+\underline{\nu}P_4\}, \hspace{5mm} {\mathcal G}_{218}=\{D+\varepsilon P_2+kP_5+\underline{\mu}P_3+\underline{\nu}Q_2\}, \\ &
{\mathcal G}_{219}=\{D+\varepsilon P_5+\underline{\mu}P_3+\underline{\nu}P_4+\underline{\rho}Q_2\}, \\ & {\mathcal G}_{220}=\{D+\varepsilon P_2+kP_5+\underline{\mu}P_3+\underline{\nu}P_4+\underline{\rho}Q_2\}, \hspace{5mm}
{\mathcal G}_{221}=\{D+\varepsilon P_2+kP_5+\underline{\mu}Q_1\}, \\ & {\mathcal G}_{222}=\{D+\varepsilon P_5+\underline{\mu}P_4+\underline{\nu}Q_1\}, \hspace{5mm}
{\mathcal G}_{223}=\{D+\varepsilon P_5+\underline{\mu}Q_1+\underline{\nu}Q_2\}, \\ & {\mathcal G}_{224}=\{D+\varepsilon P_2+kP_5+\underline{\mu}P_4+\underline{\nu}Q_1\}, \hspace{5mm}
{\mathcal G}_{225}=\{D+\varepsilon P_2+kP_5+\underline{\mu}Q_1+\underline{\nu}Q_2\}, \hspace{5cm} \\ & {\mathcal G}_{226}=\{D+\varepsilon P_5+\underline{\mu}P_4+\underline{\nu}Q_1+\underline{\rho}Q_2\}, \\ &
{\mathcal G}_{227}=\{D+\varepsilon P_2+kP_5+\underline{\mu}P_4+\underline{\nu}Q_1+\underline{\rho}Q_2\}, \\ & {\mathcal G}_{228}=\{D+\varepsilon P_1+kP_2+\ell P_5+\underline{\mu}P_3\}, \hspace{5mm}
{\mathcal G}_{229}=\{D+\varepsilon P_1+kP_5+\underline{\mu}P_3+\underline{\nu}P_4\}, \\ & {\mathcal G}_{230}=\{D+\varepsilon P_1+kP_5+\underline{\mu}P_3+\underline{\nu}Q_2\}, \\ &
{\mathcal G}_{231}=\{D+\varepsilon P_1+kP_2+\ell P_5+\underline{\mu}P_3+\underline{\nu}P_4\}, \\ & {\mathcal G}_{232}=\{D+\varepsilon P_1+kP_2+\ell P_5+\underline{\mu}P_3+\underline{\nu}Q_2\}, \\ &
{\mathcal G}_{233}=\{D+\varepsilon P_1+kP_5+\underline{\mu}P_3+\underline{\nu}P_4+\underline{\rho}Q_2\}, \\ & {\mathcal G}_{234}=\{D+\varepsilon P_1+kP_2+\ell P_5+\underline{\mu}P_3+\underline{\nu}P_4+\underline{\rho}Q_2\}, \\ &
{\mathcal G}_{235}=\{D+\varepsilon P_1+kP_2+\ell P_5+\underline{\mu}Q_1\}, \hspace{5mm} {\mathcal G}_{236}=\{D+\varepsilon P_1+kP_5+\underline{\mu}P_4+\underline{\nu}Q_1\}, \\ &
{\mathcal G}_{237}=\{D+\varepsilon P_1+kP_5+\underline{\mu}Q_1+\underline{\nu}Q_2\}, \\ & {\mathcal G}_{238}=\{D+\varepsilon P_1+kP_2+\ell P_5+\underline{\mu}P_4+\underline{\nu}Q_1\}, \\ &
{\mathcal G}_{239}=\{D+\varepsilon P_1+kP_2+\ell P_5+\underline{\mu}Q_1+\underline{\nu}Q_2\}, \\ & {\mathcal G}_{240}=\{D+\varepsilon P_1+kP_5+\underline{\mu}P_4+\underline{\nu}Q_1+\underline{\rho}Q_2\}, \\ &
{\mathcal G}_{241}=\{D+\varepsilon P_1+kP_2+\ell P_5+\underline{\mu}P_4+\underline{\nu}Q_1+\underline{\rho}Q_2\}, \\ & {\mathcal G}_{242}=\{D+\varepsilon P_2+kP_5+\underline{\mu}P_3+\underline{\nu}Q_1\}, \hspace{5mm}
{\mathcal G}_{243}=\{D+\varepsilon P_5+\underline{\mu}P_3+\underline{\nu}P_4+\underline{\rho}Q_1\}, \\ & {\mathcal G}_{244}=\{D+\varepsilon P_5+\underline{\mu}P_3+\underline{\nu}Q_1+\underline{\rho}Q_2\}, \\ &
{\mathcal G}_{245}=\{D+\varepsilon P_2+kP_5+\underline{\mu}P_3+\underline{\nu}P_4+\underline{\rho}Q_1\}, \\ & {\mathcal G}_{246}=\{D+\varepsilon P_2+kP_5+\underline{\mu}P_3+\underline{\nu}Q_1+\underline{\rho}Q_2\}, \\ &
{\mathcal G}_{247}=\{D+\varepsilon P_5+\underline{\mu}P_3+\underline{\nu}P_4+\underline{\rho}Q_1+\underline{\sigma}Q_2\}, \\ & {\mathcal G}_{248}=\{D+\varepsilon P_2+kP_5+\underline{\mu}P_3+\underline{\nu}P_4+\underline{\rho}Q_1+\underline{\sigma}Q_2\}, \\ &
{\mathcal G}_{249}=\{D+\varepsilon P_1+kP_2+\ell P_5+\underline{\mu}P_3+\underline{\nu}Q_1\}, \\ & {\mathcal G}_{250}=\{D+\varepsilon P_1+kP_5+\underline{\mu}P_3+\underline{\nu}P_4+\underline{\rho}Q_1\}, \\ &
{\mathcal G}_{251}=\{D+\varepsilon P_1+kP_5+\underline{\mu}P_3+\underline{\nu}Q_1+\underline{\rho}Q_2\},
\end{split}
\end{displaymath}
\begin{equation}
\begin{split}
&{\mathcal G}_{252}=\{D+\varepsilon P_1+kP_2+\ell P_5+\underline{\mu}P_3+\underline{\nu}P_4+\underline{\rho}Q_1\}, \\ &
{\mathcal G}_{253}=\{D+\varepsilon P_1+kP_2+\ell P_5+\underline{\mu}P_3+\underline{\nu}Q_1+\underline{\rho}Q_2\}, \\ & {\mathcal G}_{254}=\{D+\varepsilon P_1+kP_5+\underline{\mu}P_3+\underline{\nu}P_4+\underline{\rho}Q_1+\underline{\sigma}Q_2\},  \\ &
{\mathcal G}_{255}=\{D+\varepsilon P_1+kP_2+\ell P_5+\underline{\mu}P_3+\underline{\nu}P_4+\underline{\rho}Q_1+\underline{\sigma}Q_2\}. \hspace{4cm}
\end{split}
\label{list5}
\end{equation}

\noindent Therefore, the classification includes the 255 non-equivalent subalgebras listed above. It should be noted that we obtain far more subalgebras for this supersymmetric extension than were obtained for the scalar Born-Infeld supersymmetric extension described in \cite{Hariton}.

\noindent It is worth noting that the minimal surface equation (\ref{eqmotion2}) is invariant under the discrete reflection transformation
\begin{equation}
 x\rightarrow y,\quad y\rightarrow x,\quad \theta_1\rightarrow \theta_2,\quad \theta_2\rightarrow \theta_1.
\label{discreter}
\end{equation}
Hence, identifying each subalgebra of the classification with its partner equivalent under the transformation (\ref{discreter}), the subalgebra classification of ${\mathcal G}$ can be simplified from 255 to 143 subalgebras. These 143 subalgebras, labeled  ${\mathcal L}_1$ to ${\mathcal L}_{143}$, are listed in the Appendix.

\section{Symmetry group reductions and solutions of the SUSY minimal surface equation}

Each subalgebra given in the Appendix can be used to perform a symmetry reduction of the supersymmetric minimal surface equation (\ref{eqmotion2}) which, in most cases, allows us to determine invariant solutions of the SUSY MS equation. Once a solution of the equation is known, new solutions can be found by acting on the given solution with the supergroup of symmetries. Since both the equation and the list of subalgebras are very involved, we do not consider all possible cases. Instead we present certain interesting examples of nontrivial solutions which illustrate the symmetry reduction method. In each case, we begin by constructing a complete set of invariants (functions which are preserved by the symmetry subgroup action). Next, we find the group orbits of the corresponding subgroups as well as the associated reduced systems of equations. Each reduced system can be solved in order to construct an invariant solution of the SUSY MS equation (\ref{eqmotion2}). It should be noted that, as has been observed for other similar supersymmetric extensions \cite{Polytropic}, some of the subalgebras listed in the Appendix have a non-standard invariant structure in the sense that they do not reduce the system to ordinary differential equations in the usual sense. These are the 9 subalgebras: ${\mathcal L}_{2}$, ${\mathcal L}_{3}$, ${\mathcal L}_{6}$, ${\mathcal L}_{15}$, ${\mathcal L}_{16}$, ${\mathcal L}_{19}$, ${\mathcal L}_{21}$, ${\mathcal L}_{24}$, ${\mathcal L}_{33}$. This leaves 134 subalgebras that lead to standard symmetry reductions, of which we illustrate several examples.

\subsection{Translation-invariant solutions}

We first construct the following three polynomial translation-invariant solutions. For each of these examples, $K$, $C_1$, $C_2$, $C_7$ and $C_8$ are bosonic constants, while $\underline{C}$, $\underline{C_3}$, $\underline{C_4}$, $\underline{C_5}$ and $\underline{C_6}$ are fermionic constants.

\noindent{\bf 1.} For the subalgebra ${\mathcal L}_1=\{\partial_x\}$, the set of invariants is $y$, $\theta_1$, $\theta_2$, $\Phi$, which leads to the group orbit $\Phi=\Phi(y,\theta_1,\theta_2)$. Substituting into equation (\ref{eqmotion2}), we obtain the quadratic solution
\begin{equation}
\Phi(y,\theta_1,\theta_2)=C_1+C_2y+\underline{C_3}\theta_1+\underline{C_4}y\theta_1+\underline{C_5}\theta_2+\underline{C_6}y\theta_2+C_7\theta_1\theta_2+C_8y\theta_1\theta_2.
\label{solutionG1}
\end{equation}

\noindent{\bf 2.} For the subalgebra ${\mathcal L}_4=\{\partial_x+\underline{\mu}\partial_{\theta_1}\}$, we obtain the invariants $y$, $\eta=\theta_1-\underline{\mu}x$, $\theta_2$, $\Phi$, so $\Phi=\Phi(y,\eta,\theta_2)$ is the group orbit and we get the translationally invariant solution
\begin{equation}
\begin{split}
\Phi(x,y,\theta_1,\theta_2)=&C_1+C_2y+\underline{C_3}(\theta_1-\underline{\mu}x)+\underline{C_4}y(\theta_1-\underline{\mu}x)+\underline{C_5}\theta_2+\underline{C_6}y\theta_2\\ &+C_7(\theta_1-\underline{\mu}x)\theta_2+C_8y(\theta_1-\underline{\mu}x)\theta_2.
\label{solutionG4}
\end{split}
\end{equation}
which constitutes an analogue travelling wave involving both the bosonic variable $x$ and the fermionic variable $\theta_1$. Along any curve $\theta_1-\underline{\mu}x=\underline{C}$, solution (\ref{solutionG4}) depends only on $y$ and $\theta_2$, which constitutes a subcase of (\ref{solutionG1}).

\noindent{\bf 3.} The subalgebra ${\mathcal L}_{8}=\{\partial_x+k\partial_{y}\}$, $k\neq 0$, has invariants $\xi=y-kx$, $\theta_1$, $\theta_2$, $\Phi$, so we have the group orbit $\Phi=\Phi(\xi,\theta_1,\theta_2)$. We obtain the following stationary wave solution
\begin{equation}
\begin{split}
\Phi(x,y,\theta_1,\theta_2)=&C_1+C_2(y-kx)+\underline{C_3}\theta_1+\underline{C_4}\theta_1(y-kx)+\underline{C_5}\theta_2+\underline{C_6}\theta_2(y-kx)\\ &+C_7\theta_1\theta_2+C_8\theta_1\theta_2(y-kx).
\label{solutionG15}
\end{split}
\end{equation}
which is an analogue travelling wave involving the bosonic spatial variables $x$ and $y$. Along any straight line $y-kx=K$, the dependence of solution (\ref{solutionG15}) is purely fermionic.

\subsection{Scaling-invariant solution}

We first present two subalgebra reductions involving combinations of dilations and translations.

\noindent {\bf 4.} The subalgebra ${\mathcal L}_{74}=\{2x\partial_x+2y\partial_y+(\theta_1+\underline{\mu})\partial_{\theta_1}+\theta_2\partial_{\theta_2}+4\Phi\partial_{\Phi}\}$ involves a linear combination of the dilation $D$ and the fermionic translation $P_3$. This subalgebra has invariants
\begin{equation}
\xi=\dfrac{y}{x},\quad \eta_1=\dfrac{\theta_1+\underline{\mu}}{\sqrt{x}},\quad \eta_2=\dfrac{\theta_2}{\sqrt{x}},\quad \Psi=\dfrac{\Phi}{x^2},
\end{equation}
which leads to the group orbit $\Phi=x^2\Psi(\xi,\eta_1,\eta_2)$. If we make the assumption that the bosonic superfield $\Psi$ is of the particular bodiless form
\begin{equation}
\Psi=\omega(\xi)\eta_1\eta_2,
\label{bodiless}
\end{equation}
where $\omega(\xi)$ is an arbitrary bosonic function of $\xi$, equation (\ref{eqmotion2}) reduces to the ordinary differential equation
\begin{equation}
(\omega^2+\xi^2+1)\omega_{\xi\xi}\eta_1\eta_2=0,
\end{equation}
from where we obtain the following two solutions for $\omega$:
\begin{equation}
\omega(\xi)=\varepsilon_1 i\sqrt{\xi^2+1},\hspace{2cm} \omega(\xi)=A\xi+B,
\end{equation}
where $\varepsilon_1=\pm 1$ and $A$ and $B$ are complex-valued constants. This leads to the following radical and algebraic invariant solutions, respectively:
\begin{equation}
\mbox{(i) }\hspace{2mm} \Phi=\varepsilon_1 i\sqrt{x^2+y^2}(\theta_1+\underline{\mu})\theta_2,\hspace{1.5cm} \mbox{(ii) }\hspace{2mm} \Phi=(Ay+Bx)(\theta_1+\underline{\mu})\theta_2.
\label{theg66solution}
\end{equation}
Solutions (\ref{theg66solution}) consist of (i) a radially dependent solution and (ii) a centered wave whose level curves are lines intersecting at the origin. Both solutions involve the fermionic variables $\theta_1$ and $\theta_2$.

\noindent {\bf 5.} The subalgebra ${\mathcal G}_{136}=\{2x\partial_x+(2y+\varepsilon)\partial_y+\theta_1\partial_{\theta_1}+\theta_2\partial_{\theta_2}+4\Phi\partial_{\Phi}\}$ (which is equivalent to ${\mathcal L}_{73}$ under the discrete transformation (\ref{discreter})) involves a linear combination of the dilation $D$ and the bosonic translation $P_2$. This subalgebra has invariants
\begin{equation}
\xi=\dfrac{2y+\varepsilon}{x},\quad \eta_1=\dfrac{\theta_1}{\sqrt{x}},\quad \eta_2=\dfrac{\theta_2}{\sqrt{x}},\quad \Psi=\dfrac{\Phi}{x^2},
\end{equation}
which leads to the group orbit $\Phi=x^2\Psi(\xi,\eta_1,\eta_2)$. Under the assumption (\ref{bodiless}),
equation (\ref{eqmotion2}) reduces to the ordinary differential equation
\begin{equation}
(2\xi\omega\omega_{\xi}+6\omega^2+\xi^2+4)\omega_{\xi\xi}\eta_1\eta_2=0,
\end{equation}
from where we obtain the following two solutions for $\omega$:
\begin{equation}
\omega(\xi)=\varepsilon_1 \sqrt{-\dfrac{1}{8}\xi^2-\dfrac{2}{3}+\dfrac{K}{\xi^6}},\hspace{2cm} \omega(\xi)=A\xi+B,
\end{equation}
where $\varepsilon_1=\pm 1$ and $A$, $B$ and $K$ are complex-valued constants. This leads to the following radical and algebraic invariant solutions, respectively:
\begin{equation}
\mbox{(i) }\hspace{2mm} \Phi=\varepsilon_1 \theta_1\theta_2\sqrt{-\dfrac{(2y+­\varepsilon)^2}{8}-\dfrac{2x^2}{3}+\dfrac{Kx^8}{(2y+\varepsilon)^6}},\hspace{1.5cm} \mbox{(ii) }\hspace{2mm}  \Phi=\theta_1\theta_2(2Ay+Bx+\varepsilon A).
\label{theg72solution}
\end{equation}
In (\ref{theg72solution}), solution (i) is a radical solution which admits two sixth-order poles in $y$ for $\varepsilon=\pm 1$. In contrast, solution (ii) is a cubic polynomial solution which does not have poles. It is a subcase of (\ref{solutionG15}).

\noindent {\bf 6.} We now construct a scaling-invariant solution corresponding to the subalgebra ${\mathcal L}_{72}=\{2x\partial_x+2y\partial_y+\theta_1\partial_{\theta_1}+\theta_2\partial_{\theta_2}+4\Phi\partial_{\Phi}\}$. This subalgebra has invariants
\begin{equation}
\xi=\dfrac{y}{x},\quad \eta_1=\dfrac{\theta_1}{\sqrt{x}},\quad \eta_2=\dfrac{\theta_2}{\sqrt{x}},\quad \Psi=\dfrac{\Phi}{x^2},
\end{equation}
which leads to the group orbit $\Phi=x^2\Psi(\xi,\eta_1,\eta_2)$. Since the general case is very involved, we make various assumptions concerning the form of the bosonic function $\Psi$ in order to obtain particular solutions. From the hypothesis
\begin{equation}
\Psi=f(\xi)+g(\eta_1,\eta_2)+\underline{A}\eta_1+\underline{B}\eta_2+C,
\end{equation}
where $f$ and $g$ are bosonic functions, $\underline{A}$ and $\underline{B}$ are fermionic constants, and $C$ is a bosonic constant, we obtain the solutions
\begin{equation}
\Phi(x,y,\theta_1,\theta_2)=c_1x\theta_1\theta_2+c_2y^2+C_3xy+C_4x^2,
\label{solutionG64a}
\end{equation}
where $C_1$, $C_2$, $C_3$ and $C_4$ are arbitrary bosonic constants, and
\begin{equation}
\Phi(x,y,\theta_1,\theta_2)=ay^2+Cxy-Mx^2+N\theta_1\theta_2,
\label{solutionG64b}
\end{equation}
where $C$, $M$ and $N$ are arbitrary bosonic constants. Under the assumption (\ref{bodiless}),
we obtain the following double periodic solution of equation (\ref{eqmotion2})
\begin{equation}
\begin{split}
&\Phi(x,y,\theta_1,\theta_2)=\dfrac{i\theta_1\theta_2\sqrt{x}}{x^2+1}\bigg{[}2\left(-i(x+i)\right)^{1/2}2^{1/2}\left(-i(-x+i)\right)^{1/2}(xi)^{1/2}\cdot\\
&\left(x(x^2+1)\right)^{1/2}\mbox{E}\left(\left(-i(x+i)\right)^{1/2},2^{-1/2}\right)
-\left(-i(x+i)\right)^{1/2}2^{1/2}\left(-i(-x+i)\right)^{1/2}\cdot\\
&(xi)^{1/2}\left(x(x^2+1)\right)^{1/2}\mbox{F}\left(\left(-i(x+i)\right)^{1/2},2^{-1/2}\right)
-(x^3+x)^{1/2}x^2-(x^3+x)^{1/2}\bigg{]},
\end{split}
\label{solutionG64c}
\end{equation}
where $F(\varphi,k)$ and $E(\varphi,k)$ are the standard elliptic integrals of the first and second kind respectively, 
\begin{equation}
\begin{split}
&F(\varphi,k)=\int\limits_0^{\varphi}{d\theta\over \sqrt{1-k^2\sin^2{\theta}}}=\int\limits_0^{x}{dt\over \sqrt{(1-t^2)(1-k^2t^2)}},\\
&E(\varphi,k)=\int\limits_0^{\varphi}\sqrt{1-k^2\sin^2{\theta}}d\theta=\int\limits_0^{x}\sqrt{1-k^2t^2\over 1-t^2}d\theta,
\end{split}
\label{ellipticintegrals}
\end{equation}
where $x=\sin{\varphi}$, and the modulus $k=2^{-1/2}$ is such that $k^2<1$. This ensures that the elliptic solutions each possess one real and one purely imaginary period and that for real-valued arguments $\varphi$ we have real-valued solutions \cite{Byrd}. The solutions are doubly periodic multiwaves.


\noindent It should be noted that the solutions found for the subalgebras ${\mathcal L}_{74}$ and ${\mathcal G}_{136}$ involving combinations of dilations and translations were fundamentally different from the solutions found for the subalgebra ${\mathcal L}_{72}$ involving a dilation alone. It should also be noted that, at the limit where $\theta_1$ and $\theta_2$ approach zero, the solutions (\ref{theg66solution}), (\ref{theg72solution}) and (\ref{solutionG64c}) vanish. These solutions therefore have no counterpart for the classical MS equation.



\section{Group Analysis of the Classical Minimal Surface Equation}

In this section, we review previous group-theoretical results concerning the classical minimal surface equation (\ref{minimals}). In reference \cite{Bila}, the infinitesimal Lie point symmetries of (\ref{minimals}) were determined to be
\begin{equation}
\begin{split}
& e_1=\partial_x,\hspace{1cm}e_2=\partial_y,\hspace{1cm}e_3=\partial_u,\hspace{1cm}e_4=-y\partial_x+x\partial_y, \\ &  e_5=-u\partial_y+y\partial_u,\hspace{1cm} \hspace{1cm} e_6=-x\partial_u+u\partial_x,\hspace{1cm}e_7=x\partial_x+y\partial_y+u\partial_u.
\end{split}
\label{symmetriesclassical}
\end{equation}
The non-zero commutation relations of the generators (\ref{symmetriesclassical}) are given by
\begin{equation}
\begin{split}
& [e_1,e_4]=e_2,\hspace{1cm}[e_1,e_6]=-e_3,\hspace{1cm}[e_1,e_7]=e_1,\hspace{1cm}[e_2,e_4]=-e_1,\hspace{1cm} \\ & [e_2,e_5]=e_3,\hspace{1cm}[e_2,e_7]=e_2,\hspace{1cm}[e_3,e_5]=-e_2,\hspace{1cm}[e_3,e_6]=e_1,\hspace{1cm} \\ & [e_3,e_7]=e_3,\hspace{1cm}[e_4,e_5]=-e_6,\hspace{1cm}[e_4,e_6]=e_5,\hspace{1cm}[e_5,e_6]=-e_4.
\end{split}
\label{commutationclassical}
\end{equation}
The seven-dimensional Lie algebra ${\mathcal E}$ generated by the vector fields (\ref{symmetriesclassical}) can be decomposed as the following combination of semi-direct sums:
\begin{equation}
{\mathcal E}=\{\{e_4,e_5,e_6\}\sdir\{e_1,e_2,e_3\}\}\sdir\{e_7\}
\label{classicaldecomposition}
\end{equation}
Using the methods described in section 4, we perform a classification of the one-dimensional subalgebras of the Lie algebra ${\mathcal E}$. We briefly summarize the obtained results. We begin with the subalgebra ${\mathcal A}=\{e_4,e_5,e_6\}$. This subalgebra is isomorphic to $A_{3,9}\hspace{5mm} (su(2))$ as listed in \cite{Patera77}, whose subalgebras are all conjugate with $\{e_4\}$. Next, we use the methods of splitting and non-splitting subalgebras to determine the subalgebras of
\begin{equation}
{\mathcal A}\sdir{\mathcal B}=\{e_4,e_5,e_6\}\sdir\{e_1,e_2,e_3\}
\label{classicalaplusb}
\end{equation}
Through the Baker-Campbell-Hausdorff formula (\ref{BCHformula}), we find that all subalgebras of ${\mathcal A}\sdir{\mathcal B}$ are conjugate to an element of the list $\{e_1\},  \{e_4\},  \{e_4+me_3\}$, where $m$ is any real number. Finally, if we consider the full Lie algebra ${\mathcal E}$, we also obtain the subalgebra $\{e_7\}$. Thus, the full classification of the one-dimensional subalgebras of ${\mathcal E}$
\begin{equation}
\{e_1\}, \hspace{5mm} \{e_4\}, \hspace{5mm} \{e_4+me_3\}, \hspace{5mm} \{e_7\}.\hspace{5mm}
\label{classificlist}
\end{equation}
This result is different from the one obtained in \cite{Bila}, whose twelve different conjugation classes were obtained for the classification.

\noindent We perform symmetry reduction of the classical minimal surface equation for each of the four one-dimensional subalgebras given in (\ref{classificlist}). The results for subalgebras $\{e_1\}$ and $\{e_7\}$ are the same as those found in \cite{Bila}, that is planar solutions.  However, for subalgebras $\{e_4\}$ and $\{e_4+me_3\}$, we obtain the following results which are not given in \cite{Bila}. In both cases, the reduced equations led to instances of Abel's equation of the first kind.

\noindent For subalgebra $\{e_4\}$, the invariants are $\xi=x^2+y^2$ and $u$, and so $u$ is a function of $\xi$ only. Equation (\ref{minimals}) then reduces to
\begin{equation}
v_{\xi}=-\dfrac{1}{\xi}v-2v^3,\hspace{5mm} v=u_{\xi}. 
\label{thesolutionfore4}
\end{equation}
Solving equation (\ref{thesolutionfore4}) leads to the invariant solution
\begin{equation}
u(x,y)=\dfrac{1}{\sqrt{2s_0}}\ln{\Big{|}4\sqrt{s_0}\sqrt{s_0(x^2+y^2)^2-2(x^2+y^2)}+4s_0(x^2+y^2)-4\Big{|}}+k_0
\label{e4solution}
\end{equation}
where $s_0$ and $k_0$ are real constants.

\noindent For subalgebra $\{e_4+me_3\}$, the invariants are
\begin{displaymath}
\xi=x^2+y^2 \hspace{1cm}\mbox{ and }\hspace{1cm} \phi=u+m\hspace{1mm}\arcsin{\left(\dfrac{x}{\sqrt{x^2+y^2}}\right)}
\end{displaymath}
Therefore
\begin{displaymath}
u=\phi(\xi)-m\hspace{1mm}\arcsin{\left(\dfrac{x}{\sqrt{x^2+y^2}}\right)}.
\end{displaymath}
Equation (\ref{minimals}) then reduces to
\begin{equation}
v_{\xi}=-\dfrac{2\xi}{\xi+m^2}v^3-\dfrac{2\xi+3m^2}{2\xi(\xi+m^2)}v,\hspace{5mm} v=\phi_{\xi}. 
\label{thesolutionfore4plusme3}
\end{equation}
Solving equation (\ref{thesolutionfore4plusme3}) leads to the invariant solution
\begin{equation}
\begin{split}
u(x,y)=&\dfrac{im}{2}\ln{\Bigg{|}\dfrac{2\sqrt{2}im(s_0\xi-2)^{1/2}(m^2+\xi)^{1/2}+(s_0m^2-2)\xi-4m^2}{\xi}\Bigg{|}}\\ &+\dfrac{1}{\sqrt{2s_0}}\ln{\Bigg{|}\dfrac{2\sqrt{s_0}(s_0\xi-2)^{1/2}(m^2+\xi)^{1/2}+(2\xi+m^2)s_0-2}{\sqrt{s_0}}\Bigg{|}}+k_0
\end{split}
\label{e4me3solution}
\end{equation}
where $s_0$ and $k_0$ are real constants. This completes the symmetry reduction analysis of the classical minimal surface equation (\ref{minimals}) for one-dimensional Lie subalgebras.

\section{Final Remarks}

In this paper, we have formulated a supersymmetric extension of the minimal surface equation using a superspace involving two fermionic Grassmann variables and a bosonic-valued superfield. A Lie superalgebra of symmetries was determined which included translations and a dilation. The one-dimensional subalgebras of this superalgebra were classified into a large number of conjugation classes under the action of the corresponding supergroup. A number of these subalgebras were found to possess a non-standard invariant structure. For certain subalgebras, the symmetry reduction method was used to obtain invariant solutions of the SUSY MS equation. These solutions include algebraic solutions, radical solutions and doubly periodic multiwave solutions expressed in terms of elliptic integrals. In addition, we have also performed a Lie symmetry analysis of the classical minimal surface equation and compared the results with those obtained in \cite{Bila}. We found fewer one-dimensional subalgebras in the subalgebra classification by conjugation classes than obtained in \cite{Bila}. Finally, we have completed the symmetry reduction analysis for this equation. In contrast with the supersymmetric case, where 143 representative subalgebras were found, only four such subalgebras were found for the classical case. In both the classical and supersymmetric cases, a dilation symmetry was found, together with translations in all independent and dependent variables.\\\\
In the future, it would be interesting to expand our analysis in several directions. One such possibility would be to apply the above supersymmetric extension methods to the MS equation in higher dimensions. Due to the complexity of the calculations involved, this would require the development of a computer Lie symmetry package capable of handling odd and even Grassmann variables. To the best of our knowledge, such a package does not presently exist. The conservation law is well-established for the classical minimal surface equation (\ref{CL}). The question of which quantities are conserved by the supersymmetric model still remains an open question for the minimal surface equation. We could also consider conditional symmetries of the SUSY MS equation, which could allow us to enlarge the class of solutions and corresponding surfaces. Finally, it would be of interest to develop the theory of boundary conditions for equations involving Grassmann variables and analyze the existence and unicity of solutions.\\\\

\noindent {\bf Acknowledgements}\\
AMG's work was supported by a research grant from NSERC of Canada. AJH wishes to thank the Mathematical Physics Laboratory of the Centre de Recherches Math\'{e}matiques, Universit\'{e} de Montr\'{e}al, for the opportunity to participate in this research.



{}

\newpage

\begin{appendix}
\section*{Appendix}


\noindent Each subalgebra ${\mathcal G}_{1}$ to ${\mathcal G}_{255}$ listed in (\ref{list1}), (\ref{list2}), (\ref{list3}), (\ref{list4}) and (\ref{list5}) is identified with the subalgebra obtained when it is transformed using the discrete transformation (\ref{discreter}). This results in the following list of 143 conjugacy classes of one-dimensional subalgebras of the superalgebra ${\mathcal G}$ generated by the vector fields (\ref{symmetries})
\begin{displaymath}
\begin{split}
&{\mathcal L}_1=\{P_1\}, \hspace{5mm} {\mathcal L}_2=\{\underline{\mu}P_3\}, \hspace{5mm}
{\mathcal L}_3=\{\underline{\mu}Q_1\}, \hspace{5mm}  {\mathcal L}_4=\{P_1+\underline{\mu}P_3\},  \hspace{5mm}
{\mathcal L}_5=\{P_1+\underline{\mu}Q_1\}, \\ &{\mathcal L}_6=\{\underline{\mu}P_3+\underline{\nu}Q_1\},  \hspace{5mm}
{\mathcal L}_7=\{P_1+\underline{\mu}P_3+\underline{\nu}Q_1\}, \hspace{5mm} {\mathcal L}_{8}=\{P_1+kP_2\},  \hspace{5mm}
{\mathcal L}_{9}=\{P_1+\underline{\mu}P_4\}, \\ &{\mathcal L}_{10}=\{P_1+\underline{\mu}Q_2\}, \hspace{5mm}
{\mathcal L}_{11}=\{P_1+kP_2+\underline{\mu}P_4\},  \hspace{5mm} {\mathcal L}_{12}=\{P_1+kP_2+\underline{\mu}Q_2\}, \\
&{\mathcal L}_{13}=\{P_1+\underline{\mu}P_4+\underline{\nu}Q_2\},  \hspace{5mm} {\mathcal L}_{14}=\{P_1+kP_2+\underline{\mu}P_4+\underline{\nu}Q_2\}, \hspace{5mm}
{\mathcal L}_{15}=\{\underline{\mu}P_3+\underline{\nu}P_4\}, \\ &{\mathcal L}_{16}=\{\underline{\mu}P_3+\underline{\nu}Q_2\},  \hspace{5mm}
{\mathcal L}_{17}=\{P_2+\underline{\mu}P_3+\underline{\nu}P_4\}, \hspace{5mm} {\mathcal L}_{18}=\{P_2+\underline{\mu}P_3+\underline{\nu}Q_2\},  \\
&{\mathcal L}_{19}=\{\underline{\mu}P_3+\underline{\nu}P_4+\underline{\rho}Q_2\}, \hspace{5mm} {\mathcal L}_{20}=\{P_2+\underline{\mu}P_3+\underline{\nu}P_4+\underline{\rho}Q_2\}, \hspace{5mm}
{\mathcal L}_{21}=\{\underline{\mu}Q_1+\underline{\nu}Q_2\}, \\ &{\mathcal L}_{22}=\{P_2+\underline{\mu}P_4+\underline{\nu}Q_1\},  \hspace{5mm}
{\mathcal L}_{23}=\{P_2+\underline{\mu}Q_1+\underline{\nu}Q_2\}, \hspace{5mm} {\mathcal L}_{24}=\{\underline{\mu}P_4+\underline{\nu}Q_1+\underline{\rho}Q_2\},  \\
&{\mathcal L}_{25}=\{P_2+\underline{\mu}P_4+\underline{\nu}Q_1+\underline{\rho}Q_2\}, \hspace{5mm} {\mathcal L}_{26}=\{P_1+kP_2+\underline{\mu}P_3+\underline{\nu}P_4\},  \\
&{\mathcal L}_{27}=\{P_1+kP_2+\underline{\mu}P_3+\underline{\nu}Q_2\}, \hspace{5mm} {\mathcal L}_{28}=\{P_1+\underline{\mu}P_3+\underline{\nu}P_4+\underline{\rho}Q_2\},  \\
&{\mathcal L}_{29}=\{P_1+kP_2+\underline{\mu}P_3+\underline{\nu}P_4+\underline{\rho}Q_2\}, \hspace{5mm}  {\mathcal L}_{30}=\{P_1+kP_2+\underline{\mu}Q_1+\underline{\nu}Q_2\}, \\
&{\mathcal L}_{31}=\{P_1+\underline{\mu}P_4+\underline{\nu}Q_1+\underline{\rho}Q_2\}, \hspace{5mm} {\mathcal L}_{32}=\{P_1+kP_2+\underline{\mu}P_4+\underline{\nu}Q_1+\underline{\rho}Q_2\},  \\
&{\mathcal L}_{33}=\{\underline{\mu}P_3+\underline{\nu}P_4+\underline{\rho}Q_1+\underline{\sigma}Q_2\}, \hspace{5mm} {\mathcal L}_{34}=\{P_2+\underline{\mu}P_3+\underline{\nu}P_4+\underline{\rho}Q_1+\underline{\sigma}Q_2\},  \\
&{\mathcal L}_{35}=\{P_1+kP_2+\underline{\mu}P_3+\underline{\nu}P_4+\underline{\rho}Q_1+\underline{\sigma}Q_2\}, \hspace{5mm} {\mathcal L}_{36}=\{P_5\}, \hspace{5mm} 
{\mathcal L}_{37}=\{P_1+kP_5\}, \\ &{\mathcal L}_{38}=\{P_5+\underline{\mu}P_3\}, \hspace{5mm}
{\mathcal L}_{39}=\{P_5+\underline{\mu}Q_1\}, \hspace{5mm}  {\mathcal L}_{40}=\{P_1+kP_5+\underline{\mu}P_3\},  \\ &
{\mathcal L}_{41}=\{P_1+kP_5+\underline{\mu}Q_1\}, \hspace{5mm} {\mathcal L}_{42}=\{P_5+\underline{\mu}P_3+\underline{\nu}Q_1\},  \hspace{5mm}
{\mathcal L}_{43}=\{P_1+kP_5+\underline{\mu}P_3+\underline{\nu}Q_1\}, \\ & {\mathcal L}_{44}=\{P_1+kP_2+\ell P_5\},  \hspace{5mm}
{\mathcal L}_{45}=\{P_1+kP_5+\underline{\mu}P_4\}, \hspace{5mm} {\mathcal L}_{46}=\{P_1+kP_5+\underline{\mu}Q_2\}, \\ &
{\mathcal L}_{47}=\{P_1+kP_2+\ell P_5+\underline{\mu}P_4\},  \hspace{5mm} {\mathcal L}_{48}=\{P_1+kP_2+\ell P_5+\underline{\mu}Q_2\}, \\ &
{\mathcal L}_{49}=\{P_1+kP_5+\underline{\mu}P_4+\underline{\nu}Q_2\},  \hspace{5mm} {\mathcal L}_{50}=\{P_1+kP_2+\ell P_5+\underline{\mu}P_4+\underline{\nu}Q_2\}, \\ &
{\mathcal L}_{51}=\{P_5+\underline{\mu}P_3+\underline{\nu}P_4\}, \hspace{5mm} {\mathcal L}_{52}=\{P_5+\underline{\mu}P_3+\underline{\nu}Q_2\},  \hspace{5mm}
{\mathcal L}_{53}=\{P_2+kP_5+\underline{\mu}P_3+\underline{\nu}P_4\}, \\ & {\mathcal L}_{54}=\{P_2+kP_5+\underline{\mu}P_3+\underline{\nu}Q_2\}, \hspace{5mm} 
{\mathcal L}_{55}=\{P_5+\underline{\mu}P_3+\underline{\nu}P_4+\underline{\rho}Q_2\}, \\ & {\mathcal L}_{56}=\{P_2+kP_5+\underline{\mu}P_3+\underline{\nu}P_4+\underline{\rho}Q_2\}, \hspace{5mm}
{\mathcal L}_{57}=\{P_5+\underline{\mu}Q_1+\underline{\nu}Q_2\}, \\ & {\mathcal L}_{58}=\{P_2+kP_5+\underline{\mu}P_4+\underline{\nu}Q_1\},  \hspace{5mm}
{\mathcal L}_{59}=\{P_2+kP_5+\underline{\mu}Q_1+\underline{\nu}Q_2\}, \\ & {\mathcal L}_{60}=\{P_5+\underline{\mu}P_4+\underline{\nu}Q_1+\underline{\rho}Q_2\},  \hspace{5mm}
{\mathcal L}_{61}=\{P_2+kP_5+\underline{\mu}P_4+\underline{\nu}Q_1+\underline{\rho}Q_2\}, \\ 
\end{split}
\end{displaymath}
\begin{displaymath}
\begin{split}
& {\mathcal L}_{62}=\{P_1+kP_2+\ell P_5+\underline{\mu}P_3+\underline{\nu}P_4\},  \hspace{5mm}
{\mathcal L}_{63}=\{P_1+kP_2+\ell P_5+\underline{\mu}P_3+\underline{\nu}Q_2\}, \\ & {\mathcal L}_{64}=\{P_1+kP_5+\underline{\mu}P_3+\underline{\nu}P_4+\underline{\rho}Q_2\},  \hspace{5mm}
{\mathcal L}_{65}=\{P_1+kP_2+\ell P_5+\underline{\mu}P_3+\underline{\nu}P_4+\underline{\rho}Q_2\}, \\ &  {\mathcal L}_{66}=\{P_1+kP_2+\ell P_5+\underline{\mu}Q_1+\underline{\nu}Q_2\}, \hspace{5mm}
{\mathcal L}_{67}=\{P_1+kP_5+\underline{\mu}P_4+\underline{\nu}Q_1+\underline{\rho}Q_2\}, \\ & {\mathcal L}_{68}=\{P_1+kP_2+\ell P_5+\underline{\mu}P_4+\underline{\nu}Q_1+\underline{\rho}Q_2\},  \hspace{5mm}
{\mathcal L}_{69}=\{P_5+\underline{\mu}P_3+\underline{\nu}P_4+\underline{\rho}Q_1+\underline{\sigma}Q_2\}, \\ & {\mathcal L}_{70}=\{P_2+kP_5+\underline{\mu}P_3+\underline{\nu}P_4+\underline{\rho}Q_1+\underline{\sigma}Q_2\},  \\ &
{\mathcal L}_{71}=\{P_1+kP_2+\ell P_5+\underline{\mu}P_3+\underline{\nu}P_4+\underline{\rho}Q_1+\underline{\sigma}Q_2\}, \hspace{5mm} {\mathcal L}_{72}=\{D\},  \hspace{5mm}
{\mathcal L}_{73}=\{D+\varepsilon P_1\}, \\ &{\mathcal L}_{74}=\{D+\underline{\mu}P_3\},  \hspace{5mm}
{\mathcal L}_{75}=\{D+\underline{\mu}Q_1\}, \hspace{5mm} {\mathcal L}_{76}=\{D+\varepsilon P_1+\underline{\mu}P_3\},  \\
&{\mathcal L}_{77}=\{D+\varepsilon P_1+\underline{\mu}Q_1\}, \hspace{5mm} {\mathcal L}_{78}=\{D+\underline{\mu}P_3+\underline{\nu}Q_1\},  \hspace{5mm}
{\mathcal L}_{79}=\{D+\varepsilon P_1+\underline{\mu}P_3+\underline{\nu}Q_1\}, \\ &{\mathcal L}_{80}=\{D+\varepsilon P_1+kP_2\},  \hspace{5mm}
{\mathcal L}_{81}=\{D+\varepsilon P_1+\underline{\mu}P_4\}, \hspace{5mm} {\mathcal L}_{82}=\{D+\varepsilon P_1+\underline{\mu}Q_2\},  \\
&{\mathcal L}_{83}=\{D+\varepsilon P_1+kP_2+\underline{\mu}P_4\}, \hspace{5mm} {\mathcal L}_{84}=\{D+\varepsilon P_1+kP_2+\underline{\mu}Q_2\},  \\
&{\mathcal L}_{85}=\{D+\varepsilon P_1+\underline{\mu}P_4+\underline{\nu}Q_2\},  \hspace{5mm} {\mathcal L}_{86}=\{D+\varepsilon P_1+kP_2+\underline{\mu}P_4+\underline{\nu}Q_2\}, \\
&{\mathcal L}_{87}=\{D+\underline{\mu}P_3+\underline{\nu}P_4\}, \hspace{5mm} {\mathcal L}_{88}=\{D+\underline{\mu}P_3+\underline{\nu}Q_2\},  \hspace{5mm}
{\mathcal L}_{89}=\{D+\varepsilon P_2+\underline{\mu}P_3+\underline{\nu}P_4\}, \\ &{\mathcal L}_{90}=\{D+\varepsilon P_2+\underline{\mu}P_3+\underline{\nu}Q_2\},  \hspace{5mm}
{\mathcal L}_{91}=\{D+\underline{\mu}P_3+\underline{\nu}P_4+\underline{\rho}Q_2\}, \\ &{\mathcal L}_{92}=\{D+\varepsilon P_2+\underline{\mu}P_3+\underline{\nu}P_4+\underline{\rho}Q_2\},  \hspace{5mm}
{\mathcal L}_{93}=\{D+\underline{\mu}Q_1+\underline{\nu}Q_2\}, \\ &{\mathcal L}_{94}=\{D+\varepsilon P_2+\underline{\mu}P_4+\underline{\nu}Q_1\},  \hspace{5mm}
{\mathcal L}_{95}=\{D+\varepsilon P_2+\underline{\mu}Q_1+\underline{\nu}Q_2\}, \\ &{\mathcal L}_{96}=\{D+\underline{\mu}P_4+\underline{\nu}Q_1+\underline{\rho}Q_2\},  \hspace{5mm}
{\mathcal L}_{97}=\{D+\varepsilon P_2+\underline{\mu}P_4+\underline{\nu}Q_1+\underline{\rho}Q_2\}, \\
&{\mathcal L}_{98}=\{D+\varepsilon P_1+kP_2+\underline{\mu}P_3+\underline{\nu}P_4\},  \hspace{5mm}
{\mathcal L}_{99}=\{D+\varepsilon P_1+kP_2+\underline{\mu}P_3+\underline{\nu}Q_2\}, \\ &{\mathcal L}_{100}=\{D+\varepsilon P_1+\underline{\mu}P_3+\underline{\nu}P_4+\underline{\rho}Q_2\}, \hspace{5mm}
{\mathcal L}_{101}=\{D+\varepsilon P_1+kP_2+\underline{\mu}P_3+\underline{\nu}P_4+\underline{\rho}Q_2\}, \\ &{\mathcal L}_{102}=\{D+\varepsilon P_1+kP_2+\underline{\mu}Q_1+\underline{\nu}Q_2\},\hspace{5mm} 
{\mathcal L}_{103}=\{D+\varepsilon P_1+\underline{\mu}P_4+\underline{\nu}Q_1+\underline{\rho}Q_2\}, \\ &{\mathcal L}_{104}=\{D+\varepsilon P_1+kP_2+\underline{\mu}P_4+\underline{\nu}Q_1+\underline{\rho}Q_2\},  \hspace{5mm}
{\mathcal L}_{105}=\{D+\underline{\mu}P_3+\underline{\nu}P_4+\underline{\rho}Q_1+\underline{\sigma}Q_2\}, \\ &{\mathcal L}_{106}=\{D+\varepsilon P_2+\underline{\mu}P_3+\underline{\nu}P_4+\underline{\rho}Q_1+\underline{\sigma}Q_2\},  \\
&{\mathcal L}_{107}=\{D+\varepsilon P_1+kP_2+\underline{\mu}P_3+\underline{\nu}P_4+\underline{\rho}Q_1+\underline{\sigma}Q_2\},
\hspace{5mm} {\mathcal L}_{108}=\{D+\varepsilon P_5\}, \\ & 
{\mathcal L}_{109}=\{D+\varepsilon P_1+kP_5\}, \hspace{5mm}{\mathcal L}_{110}=\{D+\varepsilon P_5+\underline{\mu}P_3\}, \hspace{5mm}
{\mathcal L}_{111}=\{D+\varepsilon P_5+\underline{\mu}Q_1\}, \\ &  {\mathcal L}_{112}=\{D+\varepsilon P_1+kP_5+\underline{\mu}P_3\},  \hspace{5mm}
{\mathcal L}_{113}=\{D+\varepsilon P_1+kP_5+\underline{\mu}Q_1\}, \\ & {\mathcal L}_{114}=\{D+\varepsilon P_5+\underline{\mu}P_3+\underline{\nu}Q_1\},  \hspace{5mm}
{\mathcal L}_{115}=\{D+\varepsilon P_1+kP_5+\underline{\mu}P_3+\underline{\nu}Q_1\}, \\ & {\mathcal L}_{116}=\{D+\varepsilon P_1+kP_2+\ell P_5\},  \hspace{5mm}
{\mathcal L}_{117}=\{D+\varepsilon P_1+kP_5+\underline{\mu}P_4\}, \\ & {\mathcal L}_{118}=\{D+\varepsilon P_1+kP_5+\underline{\mu}Q_2\}, \hspace{5mm}
{\mathcal L}_{119}=\{D+\varepsilon P_1+kP_2+\ell P_5+\underline{\mu}P_4\},  \\ & {\mathcal L}_{120}=\{D+\varepsilon P_1+kP_2+\ell P_5+\underline{\mu}Q_2\}, \hspace{5mm}
{\mathcal L}_{121}=\{D+\varepsilon P_1+kP_5+\underline{\mu}P_4+\underline{\nu}Q_2\},  \\ & {\mathcal L}_{122}=\{D+\varepsilon P_1+kP_2+\ell P_5+\underline{\mu}P_4+\underline{\nu}Q_2\}, \hspace{5mm}
{\mathcal L}_{123}=\{D+\varepsilon P_5+\underline{\mu}P_3+\underline{\nu}P_4\}, \\
\end{split}
\end{displaymath}
\begin{displaymath}
\begin{split}
&{\mathcal L}_{124}=\{D+\varepsilon P_5+\underline{\mu}P_3+\underline{\nu}Q_2\},  \hspace{5mm}
{\mathcal L}_{125}=\{D+\varepsilon P_2+kP_5+\underline{\mu}P_3+\underline{\nu}P_4\}, \\ & {\mathcal L}_{126}=\{D+\varepsilon P_2+kP_5+\underline{\mu}P_3+\underline{\nu}Q_2\}, \hspace{5mm} 
{\mathcal L}_{127}=\{D+\varepsilon P_5+\underline{\mu}P_3+\underline{\nu}P_4+\underline{\rho}Q_2\}, \\ & {\mathcal L}_{128}=\{D+\varepsilon P_2+kP_5+\underline{\mu}P_3+\underline{\nu}P_4+\underline{\rho}Q_2\}, \hspace{5mm}
{\mathcal L}_{129}=\{D+\varepsilon P_5+\underline{\mu}Q_1+\underline{\nu}Q_2\}, \\ & {\mathcal L}_{130}=\{D+\varepsilon P_2+kP_5+\underline{\mu}P_4+\underline{\nu}Q_1\},  \hspace{5mm}
{\mathcal L}_{131}=\{D+\varepsilon P_2+kP_5+\underline{\mu}Q_1+\underline{\nu}Q_2\}, \\ & {\mathcal L}_{132}=\{D+\varepsilon P_5+\underline{\mu}P_4+\underline{\nu}Q_1+\underline{\rho}Q_2\},  \hspace{5mm}
{\mathcal L}_{133}=\{D+\varepsilon P_2+kP_5+\underline{\mu}P_4+\underline{\nu}Q_1+\underline{\rho}Q_2\}, \\ &
 {\mathcal L}_{134}=\{D+\varepsilon P_1+kP_2+\ell P_5+\underline{\mu}P_3+\underline{\nu}P_4\},  \\ &
{\mathcal L}_{135}=\{D+\varepsilon P_1+kP_2+\ell P_5+\underline{\mu}P_3+\underline{\nu}Q_2\}, \\ & {\mathcal L}_{136}=\{D+\varepsilon P_1+kP_5+\underline{\mu}P_3+\underline{\nu}P_4+\underline{\rho}Q_2\},  \\ &
{\mathcal L}_{137}=\{D+\varepsilon P_1+kP_2+\ell P_5+\underline{\mu}P_3+\underline{\nu}P_4+\underline{\rho}Q_2\}, \\ &  {\mathcal L}_{138}=\{D+\varepsilon P_1+kP_2+\ell P_5+\underline{\mu}Q_1+\underline{\nu}Q_2\}, \\ &
{\mathcal L}_{139}=\{D+\varepsilon P_1+kP_5+\underline{\mu}P_4+\underline{\nu}Q_1+\underline{\rho}Q_2\}, \\ & {\mathcal L}_{140}=\{D+\varepsilon P_1+kP_2+\ell P_5+\underline{\mu}P_4+\underline{\nu}Q_1+\underline{\rho}Q_2\},  \\ &
{\mathcal L}_{141}=\{D+\varepsilon P_5+\underline{\mu}P_3+\underline{\nu}P_4+\underline{\rho}Q_1+\underline{\sigma}Q_2\}, \\ & {\mathcal L}_{142}=\{D+\varepsilon P_2+kP_5+\underline{\mu}P_3+\underline{\nu}P_4+\underline{\rho}Q_1+\underline{\sigma}Q_2\},  \\ &
{\mathcal L}_{143}=\{D+\varepsilon P_1+kP_2+\ell P_5+\underline{\mu}P_3+\underline{\nu}P_4+\underline{\rho}Q_1+\underline{\sigma}Q_2\}, \hspace{5mm}
\end{split}
\end{displaymath}

\end{appendix}

\end{document}